\documentclass[11pt]{article}
\usepackage[pdftex]{graphicx,color} 
\usepackage{jheppub}
\usepackage{amsmath}
\usepackage{amssymb}
\usepackage{comment}
\usepackage{multirow}
\usepackage{mathtools}
\usepackage{dsdshorthand}
\usepackage{tikz}
\usetikzlibrary{arrows}
\tikzset{
>=stealth'
}
\newcommand{\bea}{\begin{equation}\begin{aligned}}
\newcommand{\eea}[1]{\label{#1}\end{aligned}\end{equation}}
\newcommand{\beq}{\begin{equation}}
\newcommand{\eeq}{\end{equation}}

\title{AdS Virasoro-Shapiro from single-valued periods}

\author{Luis F. Alday, Tobias Hansen and Joao A. Silva}

\affiliation{Mathematical Institute, University of Oxford,
Woodstock Road, Oxford, OX2 6GG, UK}

\abstract{
We determine the full $1/\sqrt{\lambda}$ correction to the flat-space Wilson coefficients which enter the AdS Virasoro-Shapiro amplitude in $\mathcal{N}=4$ SYM theory at strong coupling. The assumption that the Wilson coefficients are in the ring of single-valued multiple zeta values, as expected for closed string amplitudes, is surprisingly powerful and leads to a unique solution to the dispersive sum rules relating Wilson coefficients and OPE data obtained in \cite{Alday:2022uxp}. The corresponding OPE data fully agrees with and extends the results from integrability. The Wilson coefficients to order $1/\sqrt{\lambda}$ can be summed into an expression whose structure of poles and residues generalises that of the Virasoro-Shapiro amplitude in flat space.
}

\emailAdd{alday@maths.ox.ac.uk, tobias.hansen@maths.ox.ac.uk, joao.silva@maths.ox.ac.uk}

\begin{document}
\maketitle

\section{Introduction}

Closed string genus-0 four-point amplitudes are commonly referred to as Virasoro-Shapiro amplitudes. The worldsheet theory to compute the corresponding amplitude on AdS$_5 \times$ S$^5$ space-time is still unknown, which motivates us to explore alternative methods.
In our favour we have that, by the AdS/CFT duality, this amplitude is also a correlator in $\mathcal{N} = 4$ SYM theory.
The AdS/CFT dictionary maps the genus expansion in $g_s$ to the expansion in inverse powers of the central charge $1/c$ while $\alpha'$ corresponds to inverse powers of the t'Hooft coupling $1/\sqrt{\lambda}$ . We study the correlator of four stress-tensor multiplets in Mellin space, to leading order in $1/c$, which in a $1/\lambda$ expansion can be written as (see \cite{Alday:2022uxp} for further details)
\begin{equation}
M(s_1,s_2) = \frac{8}{(s_1-\frac23)(s_2-\frac23)(s_3-\frac23)} + \sum\limits_{a,b=0}^\infty  \frac{\Gamma(2a+3b+6)}{8^{a+b}\lambda^{\frac32 + a + \frac32 b} } \sigma_2^a \sigma_3^b
\left(  \alpha^{(0)}_{a,b} + \frac{\alpha^{(1)}_{a,b} }{\sqrt{\lambda}}  + \frac{ \alpha^{(2)}_{a,b}}{\lambda}
 + \cdots \right)
\label{M}
\end{equation}
where $s_1+s_2+s_3=0$, $\sigma_2=s_1^2+s_2^2+s_3^2$ and $\sigma_3=s_1 s_2 s_3$, and we have suppressed an overall $1/c$.\footnote{These Mellin variables are related to the ones in \cite{Alday:2022uxp} by
$s_1=s-\frac43$, $s_2=t-\frac43$, $s_3=u-\frac43$. This implies $\alpha^{(0)}_{a,b} = \alpha_{a,b}$ and $\alpha^{(1)}_{a,b} = \beta_{a,b} -  \frac{2}{3} (b+1) (6+2a+3b)  \alpha^{(0)}_{a-1,b+1}$.}
The leading coefficients $\alpha^{(0)}_{a,b}$ are known from the flat space Virasoro-Shapiro amplitude via the flat space limit formula \cite{Penedones:2010ue,Fitzpatrick:2011hu}.

In \cite{Alday:2022uxp} we used the bound on chaos \cite{Maldacena:2015waa, Caron-Huot:2021rmr} to derive dispersive sum rules that relate the Wilson coefficients $\alpha^{(k)}_{a,b}$
to the OPE data of the exchanged heavy single-trace operators with dimensions that grow as $\Delta \sim \lambda^\frac14$.
In this paper, we present a solution to these dispersive sum rules, determining the coefficients $\alpha^{(1)}_{a,b}$ and the corresponding $1/\sqrt{\lambda}$ corrections to the dimensions and structure constants of said operators.

In \cite{Alday:2022uxp} it was assumed that $\alpha^{(1)}_{a,b}$ is in the ring of multiple zeta values, has uniform transcendentality and that the sums over OPE data related to $\alpha^{(1)}_{a,b}$ are given in terms of Euler-Zagier sums. As discussed in \cite{Alday:2022uxp}, these assumptions are not enough to fully fix the $\alpha^{(1)}_{a,b}$. In the present paper we make the additional assumption that $\alpha^{(1)}_{a,b}$ is in the ring of \emph{single-valued} multiple zeta values (i.e.\ $\alpha^{(1)}_{a,b}$ are single-valued periods).
This property is known to hold for tree-level closed string amplitudes in flat space \cite{Stieberger:2013wea,Stieberger:2014hba,Schlotterer:2018zce,Brown:2019wna}, and is expected from a world-sheet perspective.  This additional assumption turns out to be surprisingly powerful and leads to a unique solution for $\alpha^{(1)}_{a,b}$. This also determines the corresponding OPE data and we give the $1/\sqrt{\lambda}$ corrections to conformal dimensions and OPE coefficients as analytic formulae for many Regge trajectories. Our solution passes several checks.
First, we reproduce the dimensions of operators on the leading Regge trajectory (including the Konishi operator) known from integrability and the two available Wilson coefficients known from localisation.
Second, we fix the solution for $\alpha^{(1)}_{a,b}$ by imposing single-valuedness for a few values of $a$. The resulting solutions turn out to be single-valued for all values of $a$ that we are able to check, in a non-trivial way. Third, our solution for $\alpha^{(1)}_{a,b}$ implies an overconstrained linear system of equations for the OPE data. That this system has a solution serves as a consistency check for $\alpha^{(1)}_{a,b}$.

Having found $\alpha^{(1)}_{a,b}$, the next step is to resum the low energy expression in \eqref{M} to obtain a simpler expression that makes the analytic structure of the amplitude manifest.
As the sum over $a$ and $b$ in \eqref{M} is divergent, we introduce the flat space transform, an integral transform equivalent to the one  in the flat space limit of \cite{Penedones:2010ue,Fitzpatrick:2011hu} but without sending the AdS radius to infinity. This is equivalent to performing a Borel resummation. For the leading term this reproduces the familiar Virasoro-Shapiro amplitude
\beq
-\frac{ \Gamma \left(- S\right) \Gamma \left(-T\right) \Gamma \left(- U \right) }{\Gamma \left(S +1\right) \Gamma \left( T+1\right) \Gamma \left( U +1\right) }
= \frac{1}{STU} + 2\sum_{a, b=0}^{\infty} \hat\sigma_2^a \hat\sigma_3^b \alpha^{(0)}_{a,b}\,,
\eeq
with $\hat\sigma_2= \frac12 ( S^2+T^2+U^2)$, $\hat\sigma_3 = STU$,
for which \cite{Zagier:2019eus} found the representation
\beq
2\sum\limits_{a,b=0}^\infty \hat\sigma_2^a \hat\sigma_3^b \alpha^{(0)}_{a,b} =
 \sum\limits_{\delta=1}^\infty \frac{1}{\delta^3} \frac{y+2}{1-x-y} \binom{z+\delta-1}{\delta-1}^2\,,
\eeq
with
$x = \hat\sigma_2/\delta^2$,
$y = \hat\sigma_3/\delta^3$,
and $z = \delta \left( \sqrt{1-4y}-1\right)/2$.
For the $1/\sqrt{\lambda}$ correction we find
\beq
2 \sum\limits_{a,b=0}^\infty \hat\sigma_2^a \hat\sigma_3^b    \alpha^{(1)}_{a,b}
= \sum\limits_{\delta=1}^\infty \sum\limits_{n=0}^{\delta-1} \frac{1}{ \delta^4} 
\mathcal{D}_n(\delta)\frac{y+2}{1-x-y} 
\binom{z+\delta-\frac{n}2-1}{\delta-n-1}^2\,,
\eeq
where $\mathcal{D}_n(\delta)$ is a third order differential operator in $x$, $y$ and $z$ which produces a crossing-symmetric expression with poles up to 4th order in $S$, $T$ and $U$.

This paper is organised as follows. In section \ref{sec:review} we review the dispersive sum rule for $\alpha^{(0)}_{a,b}$, its solution and other known data.
Section \ref{sec:solving} states the dispersive sum rule for $\alpha^{(1)}_{a,b}$, our precise assumptions and, after a short primer on single-valued multiple zeta values,
constructs $\alpha^{(1)}_{a,b}$ for $0 \leq b \leq 6$.
In section \ref{sec:spin_sums} we generalise the solutions to any value of $b$ by finding general expressions for the sums over OPE data that appear in the dispersive sum rules.
We use this to determine OPE data for many Regge trajectories.
In section \ref{sec:summing} we apply the flat space transform and resum the low energy expansion.
We conclude in section \ref{sec:conclusions}.
Appendix \ref{app:crossing-symmetric_dispersion_relations} contains a derivation of general expressions and recursion relations for the dispersive sum rules based on crossing-symmetric dispersion relations. 
In appendix \ref{app:alternative_representation} we present an alternative representation for the spin sums of section \ref{sec:spin_sums}.
Appendix \ref{app:More Bootstrap Constraints} contains an analysis of extra bootstrap constraints on the OPE data of heavy operators, other than the ones in \cite{Alday:2022uxp}.
In appendix \ref{app:leading_poles} we compute the residues of the highest order poles of the amplitude at each order in $1/\lambda$ and resum them.
In appendix \ref{app:Konishi} we summarise the state of the art of the weak and strong coupling expansions of the dimension of the Konishi operator.

\section{Review of known data}
\label{sec:review}

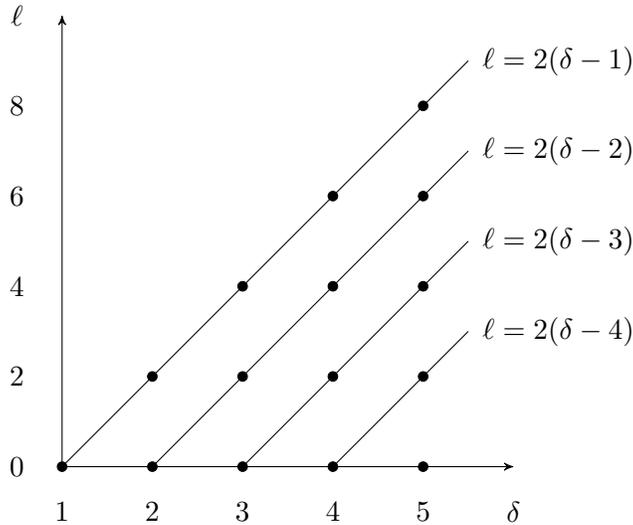
\begin{figure}[tb]
\centering
  \begin{tikzpicture}[scale=0.6]
    \coordinate (nw) at (0,10);
    \coordinate (sw) at (0,0);
    \coordinate (se) at (10,0);
    \draw[->] (sw) --  (nw) ;
    \draw[->] (sw) --  (se) ;
    \node at (0,-1) [] {$1$}; 
    \node at (2,-1) [] {$2$}; 
    \node at (4,-1) [] {$3$}; 
    \node at (6,-1) [] {$4$}; 
    \node at (8,-1) [] {$5$}; 
    \node at (10,-1) [] {$\delta$}; 
    \node at (-1,0) [] {$0$};
    \node at (-1,2) [] {$2$};
    \node at (-1,4) [] {$4$};
    \node at (-1,6) [] {$6$};
    \node at (-1,8) [] {$8$};
    \node at (-1,10) [] {$\ell$};
    \draw[-] (0,0) --  (9,9) ;
    \draw[-] (2,0) --  (9,7) ;
    \draw[-] (4,0) --  (9,5) ;
    \draw[-] (6,0) --  (9,3) ;
    \node at (11,9) [] {$\ell = 2(\delta-1)$};
    \node at (11,7) [] {$\ell = 2(\delta-2)$};
    \node at (11,5) [] {$\ell = 2(\delta-3)$};
    \node at (11,3) [] {$\ell = 2(\delta-4)$};
    \filldraw [black] (0,0) circle (3pt);
    \filldraw [black] (2,0) circle (3pt);
    \filldraw [black] (2,2) circle (3pt);
    \filldraw [black] (4,0) circle (3pt);
    \filldraw [black] (4,2) circle (3pt);
    \filldraw [black] (4,4) circle (3pt);
    \filldraw [black] (6,0) circle (3pt);
    \filldraw [black] (6,2) circle (3pt);
    \filldraw [black] (6,4) circle (3pt);
    \filldraw [black] (6,6) circle (3pt);
    \filldraw [black] (8,0) circle (3pt);
    \filldraw [black] (8,2) circle (3pt);
    \filldraw [black] (8,4) circle (3pt);
    \filldraw [black] (8,6) circle (3pt);
    \filldraw [black] (8,8) circle (3pt);
  \end{tikzpicture}
\caption{Chew-Frautschi plot of the stringy operators.} \label{fig:chew_frautschi}
\end{figure}

The `stringy' operators that enter the dispersive sum rules for $\alpha^{(k)}_{a,b}$ are the single trace operators with twists $\tau(r; \lambda)$ and OPE coefficients $\mathcal{C}^2(r; \lambda)$, which can be parameterised by
\begin{align}
\tau(r; \lambda) &= \tau_0(r) \lambda^{\frac{1}{4}} + \tau_1(r)  +  \tau_2(r) \lambda^{-\frac{1}{4}} + \ldots \,, \label{twistsStringy}\\
\mathcal{C}^2(r; \lambda) &=  \frac{\pi ^3}{2^{12}} \frac{ 2^{-2 \tau(r; \lambda)} \tau(r; \lambda)^6 }{\sin^2(\frac{\pi \tau(r; \lambda)}{2}) } \frac{1}{2^{2 \ell}(\ell+1)}  f(r; \lambda)\,, \label{OPEStringy1} \\
 f(r; \lambda) &= f_0(r) +  f_1(r) \lambda^{-\frac{1}{4}} + f_2(r) \lambda^{-\frac{1}{2}} + \ldots \,, \label{OPEStringy2}
\end{align}
where $r$ labels collectively all quantum numbers characterising the operators.
The leading contribution to the twists is
\begin{align}
 \tau_0(r) = 2\sqrt{\delta}\,, \qquad \delta \in \mathbb{N} \,,
\end{align}
and we will use $\delta$ along with the spin $\ell$ to label operators from now on, i.e.\ $r =(\delta, \ell, \hat r)$.
As long as we are studying only a single correlator we cannot access further quantum numbers $\hat r$ and will denote the sum over them by
\beq
\langle \ldots \rangle = \sum\limits_{\hat r} \ldots\,.
\eeq
The operators are organised into Regge trajectories by their dependence on $\delta$ and $\ell$ as illustrated in figure \ref{fig:chew_frautschi}.

The dispersion relations imply the following expression for the first layer of Wilson coefficients\footnote{
The derivation of the dispersive sum rules for $\alpha^{(k)}_{a,b}$ for general $b$ is described in appendix \ref{app:wilson_coefficients}.
}
\begin{equation}
\alpha^{(0)}_{a,b} = \sum\limits_{\delta=1}^\infty \sum\limits_{m=0}^b  \frac{c^{(0)}_{a,b,m}}{\delta^{3+2a+3b}} F_m^{(0)}(\delta)\,,
\label{alpha}
\end{equation}
where
\bea
c^{(0)}_{a,b,m} ={}& \frac{(-1)^m (2 a+3 b-3 m) \Gamma (a+b-m) }{2 \Gamma (a+1) \Gamma
   (b-m+1)}\\
& {}_4F_3\left(\tfrac{m+1}{2},\tfrac{m}{2},m-b,m+1-\tfrac{2
   }{3}a-b;m+1,m+1-a-b,m-\tfrac{2}{3}a-b;4\right)\,,
\eea{c0}
and the leading contributions to the OPE coefficients appear in the sums
\beq
F_m^{(0)}(\delta) = \frac{4^m}{\Gamma(2m+2)} \sum_{\ell=0,2,\ldots}^{2(\de-1)}  (\ell-m+1)_m (\ell+2)_m  \langle f_0(\de,\ell) \rangle\,.
\label{F0_def}
\eeq
The first few cases read explicitly
\bea
\alpha^{(0)}_{a,0} ={}& \sum\limits_{\de = 1}^\infty 
\frac{1}{\delta^{3+2a}} F_0^{(0)}(\delta)\,,\\
\alpha^{(0)}_{a,1} ={}& \sum\limits_{\de = 1}^\infty 
\frac{1}{\delta^{6+2a}} \left(
(a+\tfrac32) F_0^{(0)}(\delta)-F_1^{(0)}(\delta)\right)\,,\\
\alpha^{(0)}_{a,2} ={}& \sum\limits_{\de = 1}^\infty 
\frac{1}{\delta^{9+2a}} \left(
\frac{1}{2} (a+1) (a+3) F_0^{(0)}(\delta)-(a+\tfrac52) F_1^{(0)}(\delta)+F_2^{(0)}(\delta)\right)\,,\\
\vdots \ \,  &
\eea{alpha_equations}
All the coefficients $\alpha^{(0)}_{a,b}$ are known from the flat space limit, for example
\begin{align}
\alpha^{(0)}_{a,0} ={}& \zeta(3+2a)\,,\nonumber\\
\alpha^{(0)}_{a,1} ={}& \sum\limits_{\substack{i_1,i_2=0\\i_1+i_2=a}}^a\zeta(3+2i_1) \zeta(3+2i_2)
=(a+\tfrac32) \zeta (2 a+6)-2 \zeta (2 a+5,1) \,,\nonumber\\
\alpha^{(0)}_{a,2} ={}& \frac{2}{3}\sum\limits_{\substack{i_1,i_2,i_3=0\\i_1+i_2+i_3=a}}^a\zeta(3+2i_1) \zeta(3+2i_2) \zeta(3+2i_3) + \frac16 (a+1)(a+2) \zeta(9+2a) \label{alphas}\\
={}& \tfrac{1}{2} (a+1) (a+3) \zeta (2 a+9)-2(a+\tfrac52) \zeta (2 a+8,1)+ \zeta (2 a+7,2)+4 \zeta (2 a+7,1,1)\,,\nonumber\\
\vdots \ \,  &\nonumber
\end{align}
written in terms of multiple zeta values of depth $k$ and weight $s_1 + \ldots + s_k$
\beq
\zeta(s_1, \ldots, s_k) = \sum\limits_{n_1>\ldots>n_k>0} \frac{1}{n_1^{s_1} \cdots n_k^{s_k}}\,. 
\label{MZV}
\eeq
By comparing \eqref{alpha_equations} with \eqref{alphas} it is apparent that $F_m^{(0)}(\delta)$ are most naturally expressed in terms of Euler-Zagier sums, defined by
\beq
\label{nested}
Z_{s_1,\ldots,s_k} (N) = \sum\limits_{\substack{n_1,\ldots,n_k\\
N\geq n_1>\ldots>n_k>0}} \frac{1}{n_1^{s_1} \cdots n_k^{s_k}} \,, \qquad Z (N) = 1\,, \qquad Z_{s_1,\ldots,s_k} (0)=0\,,
\eeq
and which naturally lead to multiple zeta values when summed over $\delta$
\beq
\zeta(s,s_1,s_2,\ldots) = \sum\limits_{\delta=1}^\oo \frac{Z_{s_1,s_2,\ldots} (\delta-1)}{\delta^{s}} \,.
\label{ZtoZeta}
\eeq
We see that
\beq
F_m^{(0)}(\delta) = \sum\limits_{d=\lfloor \frac{m+1}{2} \rfloor}^m
\sum\limits_{\substack{s_1,\ldots,s_d \in \{1,2\}\\s_1+\ldots+s_d=m}}
2^{\sum_i \delta_{s_i,1}} \delta^m Z_{s_1,\ldots,s_d} (\delta-1)\,.
\label{F0sol}
\eeq
Equating \eqref{F0sol} and \eqref{F0_def} fixes all the individual OPE coefficients $\langle f_0(\de,\ell) \rangle$. To see this, note that $F_m^{(0)}(\delta)$ is a sum over $\delta$ different spins, so $F_m^{(0)}(\delta)$ for $0\leq m < n$ fixes all the OPE coefficients with $\delta \leq n$.

At the next order the requirement that \eqref{M} is an expansion in $1/\sqrt{\lambda}$ leads to a sum rule for vanishing Wilson coefficients
\begin{equation}
0 = \sum\limits_{\delta=1}^\infty \sum\limits_{m=0}^b  \frac{c^{(0)}_{a,b,m}}{\delta^{\frac72+2a+3b}} \left( F_m^{(1)}(\delta) - (3+2 a+3 b) T_m^{(1)}(\delta) \right)\,,
\label{zero_sum_rule}
\end{equation}
with
\bea
T_m^{(1)}(\delta) ={}& \frac{4^m}{\Gamma(2m+2)} \sum_{\ell=0,2,\ldots}^{2(\de-1)}  (\ell-m+1)_m (\ell+2)_m  
\langle f_0(\delta,\ell) (\tau_1(\delta,\ell) + \ell+2) \rangle\,,\\
F_m^{(1)}(\delta) ={}& \frac{4^m}{\Gamma(2m+2)} \sum_{\ell=0,2,\ldots}^{2(\de-1)}  (\ell-m+1)_m (\ell+2)_m 
\langle \sqrt{\delta }  f_1(\delta,\ell) -f_0(\delta,\ell) ( 3 \ell+ \tfrac{23}{4} ) \rangle\,.
\eea{T1F1_def}
This sum rule has the solution
\begin{align}
\tau_1(\delta, \ell)  = - \ell - 2\,, ~~~~~ \langle f_1( \delta, \ell) \rangle = \langle f_0(\delta, \ell) \rangle \frac{3 \ell+\frac{23}{4}}{\sqrt{\delta }}\,.
\label{tau1f1}
\end{align}
From a string theory perspective, we expect the corrections to energies of string configurations to be spaced by half-integer powers of $\lambda$
\beq
\hat\Delta(r; \lambda) = \sum\limits_{i=0}^\infty \hat\Delta_{2i}(r) \lambda^{\frac14 - \frac{i}{2}}\,.
\label{DeltaHat}
\eeq
Our result for $\tau_1(\delta, \ell)$ suggests that the states we are considering are dual to the energies from string theory by a shift of $2$ from a supersymmetry transformation
\beq
\hat\Delta(r; \lambda) = \Delta(r; \lambda) + 2 =\tau(r; \lambda)+\ell + 2 \,.
\label{Delta_to_DeltaHat}
\eeq 
For example, the dimension of the Konishi operator $\text{Tr } Z D^2 Z$ is of the form \eqref{DeltaHat} at strong coupling and $\hat\Delta_\text{classical} = 4$ in the free theory (see appendix 
\ref{app:Konishi}).
The superconformal primary of the Konishi supermultiplet $\sum_{i=1}^6 \text{Tr } \Phi_i^2$, which is exchanged in the correlator \eqref{M}, has classical dimension  $\Delta_\text{classical} = 2$, in agreement with \eqref{Delta_to_DeltaHat}.\footnote{$\Phi_i$ are the six real scalar fields of $\mathcal{N}=4$ SYM and $Z=\Phi_1+i\Phi_2$.} For this reason we expect $\tau_1(\delta, \ell)$ to be degenerate, i.e.\ the same for all species, so that $\langle  f_0(\delta, \ell) \tau_1( \delta, \ell)^2 \rangle = \langle  f_0(\delta, \ell) \rangle (\ell + 2)^2$ and so on.

Starting with the next layer of Wilson coefficients $\alpha^{(1)}_{a,b}$ and the corresponding OPE data $\langle f_0(\delta,\ell) \tau_{2}(\delta,\ell)\rangle$ and $\langle f_2(\delta,\ell) \rangle$ we are truly starting to explore the Virasoro-Shapiro amplitude in AdS. Of this data, the only pieces that were previously known are, from integrability, the twists on the leading Regge trajectory \cite{Gromov:2011de,Basso:2011rs,Gromov:2011bz}
\beq
\tau_{2}(\delta,2(\delta-1))  = \frac{3 \delta ^2-\delta +2}{2 \sqrt{\delta }}\,,
\label{integrability}
\eeq
and, from supersymmetric localisation, the Wilson coefficients\footnote{Localisation also fixes $\alpha^{(2)}_{0,0}$ and $\alpha^{(3)}_{0,0}$ as in \eqref{M_with_D8R4} below.} \cite{Binder:2019jwn,Chester:2020dja}
\beq
\alpha^{(1)}_{0,0}=0\,, \qquad
\alpha^{(1)}_{1,0} = - \frac{22}{3} \zeta(3)^2\,.
\label{localisation}
\eeq
In the remainder of the paper we will determine the rest of this data.

\section{Solving the sum rules}
\label{sec:solving}

The dispersive sum rule for the next layer of Wilson coefficients is
\bea
\alpha^{(1)}_{a,b} = \sum\limits_{\delta=1}^\infty 
  \sum\limits_{m=0}^b
\frac{1}{\delta^{4+2a+3b}} \bigg( {}&
 c^{(0)}_{a,b,m} \left( F^{(2)}_m(\delta) - (3+2a+3b)T^{(2)}_m(\delta) \right)\\
&+ c^{(2,0)}_{a,b,m} F^{(0)}_m(\delta)+ c^{(2,1)}_{a,b,m} F^{(0)}_{m+1}(\delta) \bigg)\,,
\eea{beta}
with new OPE data encoded in the sums
\begin{align}
T^{(2)}_m(\delta ) ={}& \frac{4^m}{\Gamma(2m+2)} \sum_{\ell=0,2,\ldots}^{2(\de-1)} \sqrt{\de}  (\ell-m+1)_m (\ell+2)_m  \langle f_0(\de,\ell) \tau_2(\de,\ell) \rangle\,,\label{T2}\\
F^{(2)}_m(\delta ) ={}& \frac{4^m}{\Gamma(2m+2)} \sum_{\ell=0,2,\ldots}^{2(\de-1)} (\ell-m+1)_m (\ell+2)_m \left( \de \langle f_2(\de,\ell) \rangle - \frac{39}{4} \ell \langle f_0(\de,\ell) \rangle\right)\,.
\label{F2}
\end{align}
The coefficients $c^{(0)}_{a,b,m}$ are the ones given in \eqref{c0} and the new ones are given by
\bea
c^{(2,0)}_{a,b,m} ={}& \Big(
-\left(\tfrac{1}{2}a+\tfrac{3}{4}b+\tfrac{27}{4}\right)m^2 
+ \left(2 a^2+6 a b+\tfrac{26}{3}a+\tfrac{9}{2} b^2+13 b-\tfrac{21}{8}\right)m\\
&-\tfrac{4}{3} a^3-6 a^2 b-\tfrac{16}{3} a^2-9 a b^2-16 a b-\tfrac{1}{3}a-\tfrac{9}{2} b^3-12 b^2-\tfrac{1}{2}b-\tfrac{277}{32}
\Big) c^{(0)}_{a,b,m}\,,\\
c^{(2,1)}_{a,b,m} ={}&  (m+1) \Big(
-\left(\tfrac{1}{2}a+\tfrac{3}{4}b+\tfrac{27}{4}\right)m+a^2+3 a b+\tfrac{49 }{12}a+\tfrac{9 }{4}b^2+\tfrac{49}{8}b-\tfrac{75}{16}
\Big) c^{(0)}_{a,b,m}\,.
\eea{c2}
In contrast to $\alpha^{(0)}_{a,b}$, we now have unknown data on both sides of the equations.
There are some constraints
arising from the fact that the sum rule \eqref{beta} is valid for $b=0,1,\ldots$ and $a=-b,-b+1, \ldots$ and that 
\eqref{M} needs to be an expansion in positive powers of $\sigma_2$ and $\sigma_3$, which implies
\beq
\alpha^{(1)}_{a,b} = 0 \,, \quad \text{for } a = -b, -b+1, \ldots, -1\,.
\label{locality}
\eeq
As discussed in \cite{Alday:2022uxp}, this is not enough to fully fix $\alpha^{(1)}_{a,b}$.
However, we claim that all the data is uniquely fixed by the following set of assumptions:
\begin{itemize}
\item $\alpha^{(1)}_{a,b}$ is in the ring of single-valued multiple zeta values and has uniform weight $4+2a+3b$.
\item $T^{(2)}_m(\delta)$ is a linear combination of Euler-Zagier sums (and multiple zeta values) of maximal weight $m+2$ and maximal depth $m+1$.
\item $F^{(2)}_m(\delta)$ is a linear combination of Euler-Zagier sums (and multiple zeta values) of maximal weight $m+3$ and maximal depth $m+1$.
\end{itemize}
Note that the last two assumptions imply that $\alpha^{(1)}_{a,b}$ will have multiple zeta values of maximal depth $b+2$, which is the simplest possibility compatible with the depth 2 localisation result for $\alpha^{(1)}_{1,0}$ \eqref{localisation}.

It is known that the sphere integrals of tree-level closed string amplitudes in flat space lead to Wilson coefficients in the ring of single-valued multiple zeta values.
In particular, all the $\alpha^{(0)}_{a,b}$'s can be written in terms of single zeta values of odd arguments (which are single-valued), due to the following representation for the flat space result
\bea
\frac{f(S, T)}{STU} ={}& 
-\frac{ \Gamma \left(- S\right) \Gamma \left(-T\right) \Gamma \left(- U \right) }{\Gamma \left(S +1\right) \Gamma \left( T+1\right) \Gamma \left( U +1\right) }\\
={}&
\frac{1}{STU} \exp \left(2 \sum\limits_{n=1}^{\oo}\frac{\zeta(2n+1)}{2n+1}  \left(S^{2n+1}+T^{2n+1}+U^{2n+1}\right) \right)\\
={}& \frac{1}{STU} + 2\sum_{a, b=0}^{\infty} \hat\sigma_2^a \hat\sigma_3^b \alpha^{(0)}_{a,b}\,,
\eea{flatspace_amplitude}
where $S, T, U$ are the dimensionless Mandelstam variables related to particle momenta $p_i$
\beq
\begin{gathered}
S = - \frac{\a'}{4} (p_1+p_2)^2\,, \qquad
T = - \frac{\a'}{4} (p_1+p_3)^2\,, \qquad
U = - \frac{\a'}{4} (p_1+p_4)^2\,,\\
\hat\sigma_2= \frac12 ( S^2+T^2+U^2)\,, \qquad
\hat\sigma_3 = STU\,,
\label{mandelstams}
\end{gathered}
\eeq
satisfying $S+T+U=0$. $f(S, T)$ is the four-graviton amplitude of type IIb superstring theory in flat space divided by the corresponding supergravity amplitude.
We are making the assumption that the (currently still unknown) worldsheet description of closed strings in AdS also leads to single-valued multiple zeta values.

We will concretise the assumptions on $T^{(2)}_m(\delta)$ and $F^{(2)}_m(\delta)$ and construct the solutions after a short excursion on single-valued multiple zeta values.

\subsection{(Single-valued) multiple zeta values}

In this section we give a practical introduction to working with (single-valued) multiple zeta values (MZVs).
There are many relations between MZVs of the same weight, so in order to compare MZVs, we expand them in a basis for the algebra $\mathcal{H}_N$ of MZVs of weight $N$.
We denote by $\mathcal{L}$ the space $\mathcal{H}$ modulo products of MZVs, i.e.\ 
$\mathcal{H}$ is the polynomial algebra generated by the elements of $\mathcal{L}$.
We list some examples of possible basis elements and the dimensions of both spaces and in tables \ref{LHbasis} and \ref{dimH}.
\begin{table}[h!]
\centering
\begin{tabular}{ |c|cccccccc| } 
 \hline
 $N$ & 2 & 3 & 4 & 5 & 6 & 7&8&9 \\ 
 \hline
 $\mathcal{L}_N \text{ basis}$ & $\zeta(2)$ & $\zeta(3)$ && $\zeta(5)$ && $\zeta(7)$ & $\zeta(5,3)$ & $\zeta(9)$  \\ 
 \hline
 $\mathcal{H}_N \text{ basis}$ & $\zeta(2)$ & $\zeta(3)$ &$\zeta(2)^2$  & $\zeta(5)$ & $\zeta(3)^2$ & $\zeta(7)$ & $\zeta(5,3)$ & $\zeta(9)$ \\ 
  &&&& $\zeta(3)\zeta(2)$ &$\zeta(2)^3$& $\zeta(5) \zeta(2)$ & $\zeta(5)\zeta(3)$ & $\zeta(7) \zeta(2)$ \\ 
  &&&&&& $\zeta(3) \zeta(2)^2$ & $\zeta(3)^2\zeta(2)$ & $\zeta(5) \zeta(2)^2$ \\ 
  &&&&&& & $\zeta(2)^4$ & $\zeta(3)^3$ \\ 
  &&&&&& & & $\zeta(3) \zeta(2)^3$ \\ 
 \hline
\end{tabular}
\caption{Possible basis for $\mathcal{H}$ and $\mathcal{L}$.}
\label{LHbasis}
\end{table}
\begin{table}[h!]
\centering
\begin{tabular}{ |c|cccccccccccccccccc| } 
 \hline
 $N$ & 2 & 3 & 4 & 5 & 6 & 7&8&9&10&11&12&13&14&15&16&17&18&19 \\ 
 \hline
 $\text{dim } \mathcal{L}_N$ & 1&1&0&1&0&1&1&1&1&2&2&3&3&4&5&7&8&11  \\ 
 $\text{dim } \mathcal{H}_N$ & 1&1&1&2&2&3&4&5&7&9&12&16&21&28&37&49&65&86 \\ 
 \hline
\end{tabular}
\caption{Dimensions of the space $\mathcal{H}$ ($\mathcal{L}$) of multiple zeta values of weight $N$ (modulo products).}
\label{dimH}
\end{table}
The task of rewriting MZVs in terms of a basis has been performed for weights up to 30 in \cite{Blumlein:2009cf}. In practice we use the program HyperlogProcedures \cite{HyperlogProcedures} by Oliver Schnetz for this. An example is
\beq
\zeta(3,2,1) = 3 \zeta(3)^2 - \frac{29}{30} \zeta(2)^3\,.
\eeq

Single-valued multiple zeta values were first studied by Brown \cite{Brown:2013gia} and are defined as
single-valued multiple polylogarithms evaluated at unit argument.
There is a map on the ring of multiple zeta values $\mathcal{Z}$, $\text{sv} : \mathcal{Z} \to \mathcal{Z}$
that sends multiple zeta values $\zeta(s_1,s_2,\ldots)$ to single-valued multiple zeta values $\zeta^{\text{sv}}(s_1,s_2,\ldots)$
which generate a smaller ring $\mathcal{Z}^\text{sv} \subset \mathcal{Z}$.
In particular we have
\beq
\zeta^{\text{sv}}(2k)=0\,, \qquad
\zeta^{\text{sv}}(2k+1)=2 \zeta(2k+1)\,,
\qquad k \in \mathbb{N}\,.
\eeq
The space $\mathcal{H}^\text{sv}$ of single-valued MZVs is the polynomial algebra generated by $\mathcal{L}^\text{sv}$, which is obtained by taking the elements of odd weight of $\mathcal{L}$ and applying the sv map \cite{Brown:2013gia}.
We show some basis generators and the dimensions of these spaces in tables \ref{Lsvbasis} and \ref{dimHsv}.
\begin{table}[h!]
\centering
\begin{tabular}{ |c|ccccccc| } 
 \hline
 $N$ & 3 & 5 & 7 & 9 & 11 & 13 & 15 \\ 
 \hline
 $\mathcal{L}^\text{sv}_N \text{ basis}$ & $\zeta^\text{sv}(3)$ & $\zeta^\text{sv}(5)$ & $\zeta^\text{sv}(7)$ & $\zeta^\text{sv}(9)$  & $\zeta^\text{sv}(11)$  & $\zeta^\text{sv}(13)$  & $\zeta^\text{sv}(15)$  \\ 
  & & & & & $\zeta^\text{sv}(5,3,3)$  & $\zeta^\text{sv}(7,3,3)$  & $\zeta^\text{sv}(9,3,3)$  \\ 
  & & & & & & $\zeta^\text{sv}(5,5,3)$  & $\zeta^\text{sv}(7,3,5)$  \\ 
  & & & & & &  & $\zeta^\text{sv}(6,4,3,1,1)$  \\ 
 \hline
\end{tabular}
\caption{Possible basis for $\mathcal{L}^\text{sv}$.}
\label{Lsvbasis}
\end{table}
\begin{table}[h!]
\centering
\begin{tabular}{ |c|cccccccccccccccccc| } 
 \hline
 $N$ & 2 & 3 & 4 & 5 & 6 & 7&8&9&10&11&12&13&14&15&16&17&18&19 \\ 
 \hline
 $\text{dim } \mathcal{L}_N^\text{sv}$ & 0&1&0&1&0&1&0&1&0&2&0&3&0&4&0&7&0&11  \\ 
 $\text{dim } \mathcal{H}_N^\text{sv}$ & 0&1&0&1&1&1&1&2&2&3&3&5&5&8&8&13&14&21 \\ 
 \hline
\end{tabular}
\caption{Dimensions of the space $\mathcal{H}^\text{sv}$ ($\mathcal{L}^\text{sv}$) of single-valued multiple zeta values (modulo products).}
\label{dimHsv}
\end{table}
As the sv map is implemented in HyperlogProcedures, it is easy to obtain explicit expressions for single-valued MZVs, for example
\bea
\zeta^\text{sv}(5,3) &= -10 \zeta(5) \zeta(3)\,,\\
\zeta^\text{sv}(5,3,3) &= 
2 \zeta (5,3,3)-\frac{8}{7} \zeta (5) \zeta (2)^3+\frac{12}{5} \zeta (7) \zeta
   (2)^2+90 \zeta (9) \zeta (2)-5 \zeta (3)^2 \zeta (5)\,.
\eea{sv_example}

\subsection{Construction of the solutions}

In order to solve the equation \eqref{beta} subject to our assumptions, we will construct an ansatz for $T^{(2)}_m(\delta)$ and $F^{(2)}_m(\delta)$ and fix the coefficients by imposing the vanishing of Wilson coefficients for negative values of $a$ \eqref{locality}
as well as single-valuedness for a finite number of non-negative values of $a$
\beq
\alpha^{(1)}_{a,b} \in \mathcal{Z}^\text{sv} \,, \quad \text{for } a = 0, 1,\ldots\,.
\label{singlevaluedness}
\eeq
Let us begin with $\alpha^{(1)}_{a,0}$. We make the ansatz
\begin{align}
T^{(2)}_0(\delta)={}& d_0 Z(\de-1) + d_1 \de Z_1(\de-1) +  d_2 \de^2 Z_2(\de-1)\,, \\
F^{(2)}_0(\delta)={}& c_0 Z(\de-1) + c_1 \de Z_1(\de-1) +  c_2 \de^2 Z_2(\de-1) + c_{3} \de^3 Z_{3}(\de-1)+\tl{c}_0\delta^3 \zeta(3)  Z(\delta-1)\,,\nonumber
\end{align}
which we insert together with the solution for $F^{(0)}_m(\delta)$ \eqref{F0sol} into \eqref{beta}.
As a first step we have to ensure that the sum over $\delta$ is convergent. This is the case for $a>0$, but for $a=0$ we have the divergent term
\beq
\alpha^{(1)}_{0,0} = \sum\limits_{\delta=1}^\infty \left(\frac{\tl{c}_0 \zeta(3) + c_3 Z_3 (\delta-1)}{\delta} + O(\delta^{-2}) \right)\,.
\label{beta00_leading}
\eeq
The asymptotic expansion for the Euler-Zagier sum is given by its relation to the generalised harmonic numbers
\beq
Z_s (\delta-1) = H^{(s)}_{\delta-1}= \zeta(s) - \frac{1}{(s-1) \delta^{s-1}} + O(\delta^{-s})\,, \quad s=2,3,\ldots\,,
\eeq
so we impose convergence by setting
\beq
\tl{c}_0 = -c_3\,.
\eeq
Now the sums can be done using \eqref{ZtoZeta}
\bea
\alpha^{(1)}_{a,0}={}&
-c_3 \zeta_{\text{reg}} (2 a+1) \zeta (3)
+c_3 \zeta_{\text{reg}} (2 a+1,3)
+ \left(c_2-(2 a+3) d_2\right)\zeta (2 a+2,2)\\
&+ \left(2 a^2-(2 a+3) d_1+\tfrac{49}{6}a+c_1-\tfrac{75}{8}\right)\zeta (2 a+3,1)\\
&- \left(\tfrac{4}{3} a^3+\tfrac{16}{3} a^2+(2 a+3) d_0+\tfrac{1}{3}a-c_0+\tfrac{277}{32}\right)\zeta (2 a+4)\,.
\eea{beta_a0_ansatz}
Here $\zeta_{\text{reg}}$ are shuffle-regularised multiple zeta values, as described in section 2.1 of \cite{Broedel:2014vla}. They are finite when the first argument is 1 (for instance $\zeta_{\text{reg}}(1)=0$) and agree with the usual multiple zeta values when the first argument is $2,3,\ldots$.
For each value of $a$ this expression can be rewritten in a basis of MZVs of weight $4+2a$, for instance
\bea
\alpha^{(1)}_{0,0} ={}& \frac{1}{10} \left(4 c_0+c_1+3 c_2-5 c_3-12 d_0-3 d_1-9 d_2-44\right) \zeta(2)^2\,,\\
\alpha^{(1)}_{1,0} ={}& (\ldots) \zeta(2)^3 + (\ldots) \zeta(3)^2\,,\\
\alpha^{(1)}_{2,0} ={}& (\ldots) \zeta(2)^4 +  (\ldots) \zeta(3)\zeta(5)+ (\ldots) \zeta(5,3)\,,\\
\alpha^{(1)}_{3,0} ={}& (\ldots) \zeta(2)^5 +  (\ldots) \zeta(3)\zeta(7)+ (\ldots) \zeta(5)^2 + (\ldots) \zeta(7,3)\,.
\eea{beta0_ansatz_0123}
We now demand that each expression can be written in terms of a basis of single-valued MZVs of the same weight
\bea
\alpha^{(1)}_{0,0} ={}& 0\,,\\
\alpha^{(1)}_{1,0} ={}& \kappa_1 \zeta(3)^2\,,\\
\alpha^{(1)}_{2,0} ={}& \kappa_2 \zeta(3)\zeta(5)\,,\\
\alpha^{(1)}_{3,0} ={}& \kappa_3 \zeta(3)\zeta(7)+ \kappa_4 \zeta(5)^2 \,.
\eea{beta0_sv_ansatz_0123}
Equating \eqref{beta0_ansatz_0123} and \eqref{beta0_sv_ansatz_0123} fixes all the coefficients of the ansatz
except for one parameter. The remaining parameter (along with all new parameters in the ansatz for $b=1,2$) is fixed by the constraints \eqref{locality} and \eqref{singlevaluedness} once we impose them up to $b=2$ as described below.
The solution is consistent with the localisation result \eqref{localisation}, which we use here to immediately write the fully fixed solution at $b=0$ 
\bea
d_0&=2\,, \quad d_1 = \frac14\,, \quad d_2=1\,,\\
c_0&= \frac{405}{32}\,, \quad
c_1 = \frac{89}{8}\,, \quad
c_2 = 2\,, \quad
c_3 = -2\,. 
\eea{TF0_sol}
The result for $\alpha^{(1)}_{a,0}$ is given by
\bea
\alpha^{(1)}_{a,0}={}&2 \zeta_{\text{reg}} (2 a+1)\zeta (3) -2 \zeta_{\text{reg}} (2 a+1,3)-(2 a+1) \zeta (2 a+2,2)\\
&+\frac{1}{3} \left(6 a^2+23 a+3\right) \zeta (2 a+3,1)-\frac{1}{3} \left(4 a^3+16 a^2+13 a+6\right) \zeta (2 a+4)\,,
\eea{beta_a0_result}
and is in $\mathcal{Z}^\text{sv}$ for any value of $a$, which can be shown by rewriting it in the form
\beq
\alpha^{(1)}_{a,0} = -\left(  a^2 + \frac{35}{6} a+\frac12 \right) \hspace{-2mm}\sum\limits_{\substack{i_1,i_2=0\\i_1+i_2=a-1}}^{a-1}\zeta(3+2i_1) \zeta(3+2i_2)
-2 \hspace{-2mm} \sum\limits_{\substack{i_1,i_2=0\\i_1+i_2=a-1}}^{a-1} i_1 i_2 \zeta(3+2i_1) \zeta(3+2i_2)\,.
\label{beta_a0_single_zetas}
\eeq
The reason this rewriting in terms of single zeta values is possible is that the expression has maximal depth two and the first non-trivial generators of single-valued MZVs have depth three.

We can continue in the same way for $\alpha^{(1)}_{a,1}$ by inserting the result for $T^{(2)}_0(\delta)$ and $F^{(2)}_0(\delta)$ and making an ansatz for $T^{(2)}_1(\delta)$ and $F^{(2)}_1(\delta)$, and so on.
For general $m\geq 1$ we will use the following notation for an ansatz in terms of Euler-Zagier sums with weights up to $w_\text{max}$ and depths up to $d_\text{max}$
\beq
A^i_{w_\text{max}, d_\text{max}} = \sum\limits_{w=1}^{w_\text{max}} \sum\limits_{d=1}^{d_\text{max}}
\sum\limits_{\substack{s_1,\ldots,s_d \in \mathbb{N}\\s_1+\ldots+s_d=w}}
c^i_{s_1,\ldots,s_d} \delta^w Z_{s_1,\ldots,s_d} (\delta-1)\,.
\label{ansatz_A}
\eeq
Our ansatz for $T^{(2)}_m(\delta)$ and $F^{(2)}_m(\delta)$ has the form
\begin{align}
T^{(2)}_{m\geq 1}(\delta ) 
={}& A^1_{m+2,m+1} + \delta^3 \zeta(3) A^3_{m-1,m-1}  + \delta^5 \zeta(5) A^5_{m-3,m-3}  + \delta^6 \zeta(3)^2 A^7_{m-4,m-4} + \ldots\,,\nonumber\\
F^{(2)}_{m\geq 1}(\delta ) 
={}& A^2_{m+3,m+1} + \delta^3 \zeta(3) A^4_{m,m}  + \delta^5 \zeta(5) A^6_{m-2,m-2}  + \delta^6 \zeta(3)^2 A^8_{m-3,m-3} + \ldots\,,
\label{T2F2def}
\end{align}
taking the explicit zeta values into account when determining the maximal weights and depths of the terms.
For the zeta values we include all basis elements of $\mathcal{H}^\text{sv}$ that can be multiplied with $A^i_{w_\text{max}, d_\text{max}}$ with positive $w_\text{max}$ and $d_\text{max}$. Once we fix the coefficients, only the terms with $A^1$, $A^2$ and $A^4$ will survive. At higher orders in the expansion, similar terms should produce the zeta values in the dimension of the Konishi operator \eqref{StrongCouplingAnalytics}. Interestingly, non-trivial single-valued multiple zeta values were observed to appear in the dimension of the Konishi operator at weak coupling at 8 loops \cite{Leurent:2013mr}, see \eqref{WeakCoupling}. 
The sums in \eqref{ansatz_A} start at $w=1$ to ensure $F_{m\geq 1}(1)=T_{m\geq 1}(1)=0$, which follows from the definitions \eqref{T2} and \eqref{F2}.
The number of coefficients in each ansatz is listed in table \ref{TF_coeff_count}.
\begin{table}[h!]
\centering
\begin{tabular}{ |c|cccccc| } 
 \hline
 $m$ & 0 & 1 & 2 & 3 & 4 & 5 \\ 
 \hline
 $T^{(2)}_m(\delta)$ & 3 & 6 & 15 & 33 & 70 & 145  \\ 
 $F^{(2)}_m(\delta)$ & 5 & 11 & 28 & 64 & 138 & 288 \\ 
 \hline
\end{tabular}
\caption{Number of coefficients in the ansatz for $T^{(2)}_m(\delta)$ and $F^{(2)}_m(\delta)$.}
\label{TF_coeff_count}
\end{table}

As it turns out, imposing \eqref{locality} together with \eqref{singlevaluedness} for $a=0,1,2,3$ fixes all the coefficients in each case, which we were able to show explicitly for $b=1,\ldots,5$ and for $b=6$, where we used a smaller ansatz taking into account some of the patterns we observed from the previous results. In each case the resulting expression for $\alpha^{(1)}_{a,b}$ is in $\mathcal{Z}^\text{sv}$ also for larger values of $a$, which we checked for all cases with $b\leq 6$ and weights $4+2a+3b \leq 28$.

As in the case $b=0$, the expressions for $\alpha^{(1)}_{-b,b}$ always contain $1/\delta$ terms that require cancellations in order for the sum over $\delta$ to be convergent.
For $b>0$ it turns out that one can fix all coefficients by imposing \eqref{locality} for shuffle-regularised MZVs. The solutions always lead to convergent sums in $\delta$ due to cancellations similar to \eqref{beta00_leading}.
We checked this by computing the asymptotic expansions of the Euler-Zagier sums using 
the Mathematica package HarmonicSums \cite{Ablinger:2009ovq,Ablinger:2012ufz,Ablinger:2014rba}.
For illustration we show the first few results for the sums over OPE data (all Euler-Zagier sums are evaluated at $\delta-1$)
\begin{align}
T^{(2)}_0(\delta)={}&
\delta ^2 Z_2+\tfrac14 \delta Z_1 + 2\,,\nonumber\\
T^{(2)}_1(\delta)={}&
\delta ^3 \left(Z_3+2 Z_{1,2}+3 Z_{2,1}\right)+\delta ^2 \left(\tfrac{7}{4} Z_2 + Z_{1,1}\right)+\tfrac{9}{2} \delta 
   Z_1\,,\nonumber\\
T^{(2)}_2(\delta)={}&
\delta ^4 \big(2 Z_{1,3}+3 Z_{3,1}+3 Z_{2,2}+4 Z_{1,1,2}+6 Z_{1,2,1}+8 Z_{2,1,1}\big)\nonumber\\
&+\delta ^3 \left(2 Z_3+\tfrac{15}{4} Z_{1,2}+\tfrac{23}{4} Z_{2,1}+3 Z_{1,1,1}
\right)+\delta ^2 \left(2 Z_2+9 Z_{1,1}\right)\,,\nonumber\\
T^{(2)}_3(\delta)={}&
\delta ^5 \big(Z_{2,3}+2 Z_{3,2}+4 Z_{1,1,3}+6 Z_{1,3,1}+8 Z_{3,1,1}+6 Z_{1,2,2}+7 Z_{2,1,2}+8 Z_{2,2,1}\nonumber\\
&+8 Z_{1,1,1,2}+12 Z_{1,1,2,1}+16 Z_{1,2,1,1}+20 Z_{2,1,1,1}\big)\nonumber\\
&+\delta ^4 \big(4 Z_{1,3}+6 Z_{3,1}+\tfrac{11}{2} Z_{2,2}+8 Z_{1,1,2}+12 Z_{1,2,1}+16 Z_{2,1,1}+8 Z_{1,1,1,1}\big)\nonumber\\
&+ \delta ^3 \left(\tfrac72 Z_{1,2}+\tfrac72 Z_{2,1}+16 Z_{1,1,1}\right)\,,
\label{Tm_results}
\end{align}
and
\begin{align}
F^{(2)}_0(\delta)={}& 
-2 \delta ^3  Z_3+2 \delta ^2 Z_2+\tfrac{89}{8} \delta  Z_1+\tfrac{405}{32}+2 \delta^3 \zeta(3) F^{(0)}_0(\delta)\,,\nonumber\\
F^{(2)}_1(\delta)={}& 
-2 \delta ^4 \left(2 Z_{1,3}+Z_{2,2}+3 Z_{3,1}+Z_4\right)+\delta ^3 \left(6 Z_{1,2}+11 Z_{2,1}+Z_3\right)\nonumber\\
&+\tfrac{3}{8} \delta ^2 \left(204 Z_{1,1}+59 Z_2\right)+\tfrac{761}{16}\delta  Z_1
+2 \delta^3 \zeta(3) F^{(0)}_1(\delta)\,,\nonumber\\
F^{(2)}_2(\delta)={}& 
-2 \delta ^5 \big(2 Z_{1,4}+2 Z_{2,3}+3 Z_{3,2}+3 Z_{4,1}+4 Z_{1,1,3}+2 Z_{1,2,2}+6 Z_{1,3,1}\nonumber\\
&+2 Z_{2,1,2}+3 Z_{2,2,1}+8 Z_{3,1,1}\big)+4 \delta ^4 \big(Z_{1,3}+3
   Z_{2,2}+2 Z_{3,1}+4 Z_{1,1,2}\nonumber\\
&+7 Z_{1,2,1}+11 Z_{2,1,1}-Z_4\big)+\tfrac{1}{8} \delta ^3 \left(707 Z_{1,2}+763 Z_{2,1}+2540 Z_{1,1,1}+56 Z_3\right)\nonumber\\
&+\tfrac{1}{32} \delta ^2 \left(5428 Z_{1,1}+1341 Z_2\right)
+2 \delta^3 \zeta(3) F^{(0)}_2(\delta)\,.
\end{align}
For the values of $\alpha^{(1)}_{a,1}$ and $\alpha^{(1)}_{a,2}$ we find
\begin{align}
\alpha^{(1)}_{a,1}={}&
\zeta (3) (-4 \zeta_{\text{reg}} (2 a+3,1)+(2 a+3) \zeta (2 a+4))\nonumber\\
&+2 \zeta_{\text{reg}} (2 a+3,4)+4 \zeta_{\text{reg}} (2 a+3,1,3)+6 \zeta_{\text{reg}} (2 a+3,3,1)+2 \zeta_{\text{reg}} (2 a+3,2,2)\nonumber\\
&+2 \zeta (2 a+4,3)+2 (2 a+3) \zeta (2 a+4,1,2)+(6 a+7) \zeta (2 a+4,2,1)\nonumber\\
&-\tfrac{2}{3} \left(6 a^2+25 a+15\right) \zeta (2 a+5,2)-\tfrac{8}{3} \left(3 a^2+19 a+15\right) \zeta (2 a+5,1,1)\nonumber\\
&+\tfrac{1}{6} \left(28 a^3+212 a^2+413 a+249\right) \zeta (2 a+6,1)\nonumber\\
&-\tfrac{1}{6} (2 a+3) \left(4 a^3+34 a^2+88 a+75\right) \zeta (2 a+7)\,,
\end{align}
and
\begin{align}
\alpha^{(1)}_{a,2}={}&
\zeta(3) (2 \zeta_{\text{reg}} (2 a+5,2)+8 \zeta_{\text{reg}} (2 a+5,1,1)-2 (2 a+5) \zeta (2 a+6,1)+(a+1)(a+3) \zeta (2 a+7))\nonumber\\
&-4 \zeta_{\text{reg}} (2 a+5,1,4)-6 \zeta_{\text{reg}} (2 a+5,4,1)-4 \zeta_{\text{reg}} (2 a+5,2,3)-6 \zeta_{\text{reg}} (2 a+5,3,2)\nonumber\\
&-8 \zeta_{\text{reg}} (2 a+5,1,1,3)-12 \zeta_{\text{reg}} (2 a+5,1,3,1)-16 \zeta_{\text{reg}} (2 a+5,3,1,1)\nonumber\\
&-4 \zeta_{\text{reg}} (2 a+5,1,2,2)-4  \zeta_{\text{reg}} (2 a+5,2,1,2)-6 \zeta_{\text{reg}} (2 a+5,2,2,1)\nonumber\\
&+(2 a+1) \zeta (2 a+6,4)-4 \zeta (2 a+6,1,3)-4 \zeta (2 a+6,3,1)-2 (2 a+5)  \zeta (2 a+6,2,2)\nonumber\\
&-4 (2 a+5) \zeta (2 a+6,1,1,2)-2 (6 a+13) \zeta (2 a+6,1,2,1)-4 (4 a+7) \zeta (2  a+6,2,1,1)\nonumber\\
&+(a+2) (a+3) \zeta (2 a+7,3)+\left(10 a^2+69 a+85\right) \zeta (2 a+7,1,2)\nonumber\\
&+2 (2 a+3) (3 a+14) \zeta (2 a+7,2,1)+4 \left(6 a^2+53 a+73\right) \zeta (2 a+7,1,1,1)\nonumber\\
&+\tfrac{1}{6} \left(-26 a^3-249 a^2-673 a-528\right)  \zeta (2 a+8,2)\nonumber\\
&-\tfrac{2}{3} \left(20 a^3+222 a^2+682 a+633\right) \zeta (2 a+8,1,1)\nonumber\\
&+\tfrac{1}{6} \left(22 a^4+307 a^3+1472 a^2+2945 a+2088\right)   \zeta (2 a+9,1)\nonumber\\
&-\tfrac{1}{6} (a+1) (a+3) \left(4 a^3+52 a^2+217 a+297\right) \zeta (2a+10)\,.
\label{beta12_results}
\end{align}
One can check that these expressions cannot generally be expressed in terms of single zeta values (as the $\alpha^{(0)}_{a,b}$ or $\alpha^{(1)}_{a,0}$ \eqref{beta_a0_single_zetas}), for instance
\beq
\alpha^{(1)}_{3,1} = -\frac{209279}{300} \zeta(13)-166 \zeta(3)^2 \zeta(7)-174 \zeta(3) \zeta(5)^2+\frac{2}{25} \zeta^\text{sv}(5, 5, 3)\,,
\eeq
where
\beq
\zeta^\text{sv}(5, 5, 3) = 
2 \zeta (5,5,3)+10 \zeta (5) \zeta (5,3)+20 \zeta (9) \zeta (2)^2+275 \zeta
   (11) \zeta (2)+50 \zeta (3) \zeta (5)^2\,,
\eeq
is one of the basis elements that were chosen for $\mathcal{L}^{\text{sv}}_{13}$, see table \ref{Lsvbasis}.
In the next section we will lift the results for $\alpha^{(1)}_{a,b}$, $b=0,\ldots,6$ to any value of $b$ by finding general expressions for $T^{(2)}_m(\delta)$ and $F^{(2)}_m(\delta)$.

\section{CFT data}
\label{sec:spin_sums}

\subsection{General structure}

Our solution for $\alpha^{(1)}_{a,b}$ results in some corresponding CFT-data to order $1/\sqrt{\lambda}$. At leading order a natural way to repackage the CFT data is by considering the sums over spin $F^{(0)}_m (\delta)$ \eqref{F0sol}. At this order 
it is natural to consider $T^{(2)}_m(\delta)$ and $F^{(2)}_m(\delta)$. From the results for $m=0,\ldots,6$ we see that $T^{(2)}_m(\delta)$ takes the form
\bea 
T^{(2)}_m(\delta ) 
={}& \sum\limits_{w=m}^{m+2} \sum\limits_{d=\lfloor \frac{m+1}{2} \rfloor}^{m+1}
\sum\limits_{\substack{s_1,\ldots,s_d \in \{1,2,3\}\\s_1+\ldots+s_d=w}}
t^m_{\bm{s}} \delta^w Z_{\bm{s}} (\delta-1)\,,
\eea{Tm}
i.e.\ there are Euler-Zagier sums with words $\bm{s} = s_1,\ldots,s_d$ of letters from the alphabet $\{1,2,3\}$
with weights $m$ to $m+2$ and depths $\lfloor \frac{m+1}{2} \rfloor$ to $m+1$.
We define the following function counting the number of occurrences of a given letter in a word
\beq
n^{\bm{s}}_k = \sum\limits_{i=1}^d \delta_{s_i,k}\,.
\eeq
We then find the following formula for the coefficients in \eqref{Tm}
\beq
t^{m}_{\bm{s}} = 
\begin{cases}
2^{n^{\bm{s}}_1} \left( Q^m_w (n^{\bm{s}}_1,n^{\bm{s}}_2,n^{\bm{s}}_3) + p^m_w P_{\bm{s}} \right), & n^{\bm{s}}_3 \in \{0,1\}\,,\\
0, & n^{\bm{s}}_3 > 1\,,
\end{cases}
\eeq
with coefficients
\bea
Q^m_{m+2}(n^{\bm{s}}_1,n^{\bm{s}}_2,1) &= 1\,,
\quad &&& Q^m_{m+2}(n^{\bm{s}}_1,n^{\bm{s}}_2,0) &= \frac{n^{\bm{s}}_2(n^{\bm{s}}_2+1)}{2}\,,\\
Q^m_{m+1}(n^{\bm{s}}_1,n^{\bm{s}}_2,1) &= 2\,,
\quad &&& Q^m_{m+1}(n^{\bm{s}}_1,n^{\bm{s}}_2,0) &= \frac{n^{\bm{s}}_1+n^{\bm{s}}_2 (8 n^{\bm{s}}_2+6)}{8}\,,\\
Q^m_{m}(n^{\bm{s}}_1,n^{\bm{s}}_2,1) &= 0\,,
\quad &&& Q^m_{m}(n^{\bm{s}}_1,n^{\bm{s}}_2,0) &= 2 -\frac{n^{\bm{s}}_1 (n^{\bm{s}}_1+4 n^{\bm{s}}_2-3)}{8}-\frac{n^{\bm{s}}_2(n^{\bm{s}}_2-1)}{2} \,,
\eea{q_coeffs}
and
\beq
p^m_{m+2} = \frac12\,, \qquad
p^m_{m+1} = 1\,, \qquad
p^m_{m} = 0\,.
\eeq
The object $P_{\bm{s}}$ vanishes when the word $\bm{s}$ is lexicographically ordered and is defined by
\beq
P_{\bm{s}} = \sum\limits_{i=1}^d \delta_{s_i,\max (\bm{s})} \sum\limits_{j=i+1}^d s_j (1-\delta_{s_j,\max (\bm{s})})\,.
\eeq

Next we do the same for $F^{(2)}_m(\delta)$, which takes the form
\bea 
F^{(2)}_m(\delta ) ={}& 2 \delta^3 \zeta(3)  F_m^{(0)} (\delta)
+\sum\limits_{w=m}^{m+3} \sum\limits_{d=\lfloor \frac{m+1}{2} \rfloor}^{m+1}
\sum\limits_{\substack{s_1,\ldots,s_d \in \{1,2,3,4\}\\s_1+\ldots+s_d=w}}
f^m_{\bm{s}} \delta^w Z_{\bm{s}} (\delta-1)\,.
\eea{Fm}
We find the following expression for the coefficients
\beq
f^{m}_{\bm{s}} = 
\begin{cases}
2^{n^{\bm{s}}_1} \left( R^m_w (n^{\bm{s}}_1,n^{\bm{s}}_2,n^{\bm{s}}_3,n^{\bm{s}}_4) + P^{n^{\bm{s}}_3,n^{\bm{s}}_4}_{\bm{s},w} \right), & (n^{\bm{s}}_3, n^{\bm{s}}_4) \in \{(0,1), (2,0), (1,0), (0,0)\}\,,\\
0, & \text{otherwise}\,.
\end{cases}
\eeq
The coefficients $R^m_w$ determine the coefficients $f^{m}_{\bm{s}}$ for lexicographically ordered words and are given by
\begin{align}
R^m_{m+3}(n^{\bm{s}}_1,n^{\bm{s}}_2,0,1) ={}& R^m_{m+3}(n^{\bm{s}}_1,n^{\bm{s}}_2,2,0) = -2\,,\nonumber\\
R^m_{m+2}(n^{\bm{s}}_1,n^{\bm{s}}_2,0,1) ={}& R^m_{m+2}(n^{\bm{s}}_1,n^{\bm{s}}_2,2,0) = -4\,,\nonumber\\
R^m_{m+1}(n^{\bm{s}}_1,n^{\bm{s}}_2,0,1) ={}& R^m_{m+1}(n^{\bm{s}}_1,n^{\bm{s}}_2,2,0) = 0\,,\nonumber\\
R^m_{m+3}(n^{\bm{s}}_1,n^{\bm{s}}_2,1,0) ={}&  -2 (n^{\bm{s}}_2+1)\,,\nonumber\\
R^m_{m+2}(n^{\bm{s}}_1,n^{\bm{s}}_2,1,0) ={}&  1+ n^{\bm{s}}_1-2 n^{\bm{s}}_2 \,,\label{R}\\
R^m_{m+1}(n^{\bm{s}}_1,n^{\bm{s}}_2,1,0) ={}& 2m+3 \,,\nonumber\\
R^m_{m+3}(n^{\bm{s}}_1,n^{\bm{s}}_2,0,0) ={}&  -\tfrac13 (n^{\bm{s}}_2-1) n^{\bm{s}}_2(n^{\bm{s}}_2+1)\,,\nonumber\\
R^m_{m+2}(n^{\bm{s}}_1,n^{\bm{s}}_2,0,0) ={}&  \tfrac12 (m+2)  n^{\bm{s}}_2(n^{\bm{s}}_2+1)\,,\nonumber\\
R^m_{m+1}(n^{\bm{s}}_1,n^{\bm{s}}_2,0,0) ={}&  
-\tfrac{1}{6}m^3+4 m^2+m \left((n^{\bm{s}}_2)^2+\tfrac{1}{2}n^{\bm{s}}_2+\tfrac{467}{48}\right)+\tfrac{1}{3}(n^{\bm{s}}_2)^3+(n^{\bm{s}}_2)^2+\tfrac{1}{6}n^{\bm{s}}_2+\tfrac{89}{16}\,,\nonumber\\
R^m_{m}(n^{\bm{s}}_1,n^{\bm{s}}_2,n^{\bm{s}}_3,n^{\bm{s}}_4) ={}&  
\tfrac{1}{96} \delta_{n^{\bm{s}}_3,0} \delta_{n^{\bm{s}}_4,0}
\left(-20 m^3+420 m^2-24 m n^{\bm{s}}_2+668 m+1215\right)\,.\nonumber
\end{align}
The correction terms for unordered words are
\begin{align}
2P^{0,1}_{\bm{s},m+3} ={}& 
P^{0,1}_{\bm{s},m+2} = - 2P_{\bm{s}}\,, \qquad
P^{0,1}_{\bm{s},m+1} = 0\,, \qquad
P^{n^{\bm{s}}_3,n^{\bm{s}}_4}_{\bm{s},m} = 0\,,\nonumber\\
2P^{2,0}_{\bm{s},m+3} ={}& 
P^{2,0}_{\bm{s},m+2} = 2 \sum\limits_{i=1}^d \delta_{s_i,3} (n_3^{>i}-1)(n_1^{>i} + 2 n_2^{>i})\,, \qquad 
P^{2,0}_{\bm{s},m+1} = 0\,,\nonumber\\
P^{1,0}_{\bm{s},m+3} ={}&
\sum\limits_{i=1}^d \Big(
\delta_{s_i,3} (n^{>i}_2 (-2 n^{\bm{s}}_2+n^{>i}_2-1)+n^{>i}_1
   (-n^{\bm{s}}_2+n^{>i}_2-1))+\delta_{s_i,2} n^{>i}_1 (n^{>i}_3-1)
\Big)\,,\nonumber\\
P^{1,0}_{\bm{s},m+2} ={}&
\frac{1}{2} \sum\limits_{i=1}^d \Big(
\delta_{s_i,3} \left(n^{>i}_1 (n^{\bm{s}}_1+n^{>i}_1-2 n^{\bm{s}}_2+8 n^{>i}_2+2)+2
   n^{>i}_2 (n^{\bm{s}}_1-2 n^{\bm{s}}_2+4 n^{>i}_2+5)\right)\nonumber\\
&+4   \delta_{s_i,2} n^{>i}_1 (n^{>i}_3-1)
\Big)\,,\nonumber\\
P^{1,0}_{\bm{s},m+1} ={}&
\frac{1}{4} \sum\limits_{i=1}^d 
\delta_{s_i,3} \left(n^{>i}_1 (4 n^{\bm{s}}_1+n^{>i}_1+8 n^{\bm{s}}_2+4
   n^{>i}_2+15)+4 n^{>i}_2 (2 n^{\bm{s}}_1+4
   n^{\bm{s}}_2+n^{>i}_2+8)\right)\,,\nonumber\\
P^{0,0}_{\bm{s},m+3} ={}&
\sum\limits_{i=1}^d 
\delta_{s_i,2} n^{>i}_1 (-n^{\bm{s}}_2+n^{>i}_2+1)\,,\nonumber\\
P^{0,0}_{\bm{s},m+2} ={}&
\frac{1}{2} \sum\limits_{i=1}^d 
\delta_{s_i,2} n^{>i}_1 (n^{\bm{s}}_1-2 n^{\bm{s}}_2+n^{>i}_1+8
   n^{>i}_2+5)\,,\label{PF}\\
P^{0,0}_{\bm{s},m+1} ={}&
\frac{1}{4} \sum\limits_{i=1}^d 
\delta_{s_i,2} n^{>i}_1 (4 n^{\bm{s}}_1+8 n^{\bm{s}}_2+n^{>i}_1+4
   n^{>i}_2+1)\,,
\nonumber
\end{align}
where $n_k^{>i}$ counts how many of the letters to the right of $s_i$ match $k$
\beq
n_k^{>i} = \sum\limits_{j=i+1}^d \delta_{s_j,k}\,.
\eeq
It would be interesting to study whether $T^{(2)}_m(\delta)$ and $F^{(2)}_m(\delta)$ can be understood as coming from a specific sum, similar to the expression for $F^{(0)}_m(\delta)$ that was considered in appendix A.3.2 of \cite{Green:2008uj}.
In appendix \ref{app:alternative_representation} we derive an alternative representation for $T^{(2)}_m(\delta)$ and $F^{(2)}_m(\delta)$ purely
in terms of the function $F^{(0)}_m(\delta)$.

\subsection{Twists and OPE coefficients}
\label{sec:ope_data}

By computing OPE data for many values for $\delta$ and $\ell$
we can determine analytic formulae for the twists and OPE coefficients on different Regge trajectories,
which are illustrated in figure \ref{fig:chew_frautschi}.
Recall the results for the leading OPE coefficients \cite{Alday:2022uxp}
\begin{align}
\langle f_0 (\delta, 2(\delta-1)) \rangle ={}& 
\frac{r_0(\delta)}{\delta }\,,\nonumber\\
\langle f_0 (\delta, 2(\delta-2)) \rangle ={}& 
\frac{r_1(\delta)}{3} \left(2 \delta ^2+3 \delta -8\right)\,, \label{f0_Regge}\\
\langle f_0  (\delta, 2(\delta-3)) \rangle ={}& 
\frac{r_2(\delta)}{45} \left(10 \delta ^4+43 \delta ^3+8 \delta ^2-352 \delta -192\right)\,,\nonumber
\end{align}
where
\beq
r_n(\delta) = \frac{4^{2-2 \delta } \delta ^{2 \delta -2 n-1} (2 \delta -2 n-1)}{\Gamma (\delta ) \Gamma \left(\delta -\left\lfloor \frac{n}{2}\right\rfloor \right)}\,.
\eeq
Similarly we can now find expressions for the corrections
\begin{align}
\langle f_0 \tau_{2}\rangle (\delta,2(\delta-1))  ={}&  \frac{r_0(\delta)}{2 \delta ^{3/2}}\left( 3 \delta ^2-\delta +2 \right)\,,\nonumber\\
\langle f_0 \tau_{2}\rangle (\delta,2(\delta-2))  ={}&  \frac{r_1(\delta)}{18 \sqrt{\delta }}
\left(18 \delta ^4+25 \delta ^3-57 \delta ^2+50 \delta -72\right)\,,\label{tau2_Regge}\\
\langle f_0 \tau_{2}\rangle (\delta,2(\delta-3))  ={}&  \frac{r_2(\delta)}{1350 \sqrt{\delta }}
\left(450 \delta ^6+1985 \delta ^5+1043 \delta ^4-12782 \delta ^3-2552 \delta ^2-35712 \delta -11520\right)\,,\nonumber
\end{align}
as well as
\begin{align}
\langle f_2  (\delta,2(\delta-1)) \rangle ={}&  
-\frac{r_0(\delta)}{96 \delta ^2} \left( 112 \delta ^3-1872 \delta ^2+344 \delta +201\right)
+ 2 \delta^2 \zeta(3)\langle f_0  (\delta,2(\delta-1)) \rangle\,,\nonumber\\
\langle f_2  (\delta,2(\delta-2)) \rangle ={}&  
-\frac{r_1(\delta)}{864 \delta }
\left(672 \delta ^5-8272 \delta ^4+3072 \delta ^3+77038 \delta ^2-122559 \delta +44136\right)\nonumber\\
&+ 2 \delta^2 \zeta(3)\langle f_0 (\delta,2(\delta-2))\rangle \,,\label{f2_Regge}\\
\langle f_2  (\delta,2(\delta-3)) \rangle ={}&  
-\frac{r_2(\delta)}{324000 \delta }
\big(84000 \delta ^7-554800 \delta ^6+342368 \delta ^5+14918998 \delta ^4+28865953 \delta ^3\nonumber\\
&-197973672 \delta ^2+72891360 \delta +108388800 \big)
+ 2 \delta^2 \zeta(3)\langle f_0 (\delta,2(\delta-3)) \rangle\,.\nonumber
\end{align}
We include $\langle f_0 (\delta, \ell) \rangle$, $\langle f_0 \tau_{2}\rangle (\delta, \ell)$ and $\langle f_2 (\delta, \ell) \rangle$  for the first seven Regge trajectories in a Mathematica notebook. For the leading Regge trajectory with $\ell=2(\delta-1)$ the heavy operators are supposed to be non-degenerate. Our result for $\tau_2(\delta,\ell)$ in this case agrees exactly with the integrability results! Furthermore our procedure leads to a wealth of CFT data, including structure constants of operators in the leading Regge trajectory, which can hopefully be confronted with integrability results in the near future.  

\subsection{Checks}
\label{sec:checks}

We have performed several checks that back up both the assumptions made in section \ref{sec:solving} and the general expressions \eqref{Tm} and \eqref{Fm}.
By combining \eqref{beta} with \eqref{Tm} and \eqref{Fm} we can generate explicit expressions for $\alpha^{(1)}_{a,b}$ for any $b$ which satisfy
\eqref{locality} and \eqref{singlevaluedness}, which we checked for all cases with $b\leq 12$ and weights $4+2a+3b \leq 25$.

Another check is that
the equations \eqref{T2} = \eqref{Tm} for $T^{(2)}_m(\delta)$ or  \eqref{F2} = \eqref{Fm} for $F^{(2)}_m(\delta)$ 
for a fixed value of $\delta$ depend on $\delta$ unknown bits of CFT data (for $\ell = 0,2,\ldots, 2(\delta-1)$),
but there are $2\delta-1$ equations for these unknowns (for $m=0,1,\ldots,2(\delta-1)$).
That this overconstrained system of equation has a solution is a consistency check between the solutions for $T^{(2)}_m(\delta)$ or $F^{(2)}_m(\delta)$ for different values of $m$.

Finally, our results for $\tau_2(\delta,\ell)$ on the leading Regge trajectory $\ell=2(\delta-1)$ agree precisely with the known result from integrability \eqref{integrability}, which we checked explicitly for $\delta = 1, \ldots, 28$, and we reproduce the two Wilson coefficients that were previously determined using localisation \eqref{localisation}.

\section{Summing the low energy expansion}
\label{sec:summing}

Next we would like to do the sums over $a$, $b$ and $m$ to obtain an expression for the amplitude with explicit poles.
For the flat space amplitude this was done in \cite{Zagier:2019eus} where they found essentially the formula \eqref{alpha} by studying the Virasoro-Shapiro amplitude, without any reference to CFT dispersion relations.
In AdS we have the problem that the gamma function in \eqref{M} makes the sums over $a$ and $b$ divergent.
For this reason we will use the flat space limit formula of \cite{Penedones:2010ue,Fitzpatrick:2011hu} to regulate the sums, not only for the flat space part of the amplitude but also for the $1/\sqrt{\lambda}$ corrections. In this case we call it the flat space transform.
This essentially amounts to a Borel summation as it removes the gamma function from the sums. The flat space transform is defined by
\beq \label{flat}
\text{FS}(M(s_1,s_2)) = 2 \lambda^\frac{3}{2} c \int_{\kappa-i\infty}^{\kappa+ i \infty} \frac{d\alpha}{2 \pi i} \, e^\alpha \alpha^{-6} 
M \left( \frac{2\sqrt{\lambda} S}{\alpha}, \frac{2\sqrt{\lambda} T}{\alpha} \right)  \,,
\eeq
where $S$, $T$ and $U$ are the Mandelstams introduced in \eqref{mandelstams} and $M(s_1,s_2)$ is obtained by eliminating $s_3$ using $s_1+s_2+s_3=0$.

Applying the flat space transform to our Mellin amplitude \eqref{M} we get
\bea
\text{FS}(M(s_1,s_2)) ={}& A^{(0)}(S,T) + \frac{1}{\sqrt{\lambda}} A^{(1)}(S,T) + O(1/\lambda)\,,\\
A^{(0)}(S,T) ={}& \frac{1}{STU} + 2 \sum\limits_{a,b=0}^\infty  \hat\sigma_2^a \hat\sigma_3^b    \alpha^{(0)}_{a,b}\,,\\
A^{(1)}(S,T) ={}& -\frac{2}{3} \frac{\hat\sigma_2}{\hat\sigma_3^2}+
2\sum\limits_{a,b=0}^\infty  \hat\sigma_2^a \hat\sigma_3^b   \alpha^{(1)}_{a,b} \,.
\eea{FS_expansion}
We would like to do the sum in the expression for $A^{(0)}(S,T)$.
To this end we use the following representation of the coefficients $c^{(0)}_{a,b,m}$ from \eqref{c0}
\begin{equation}
c^{(0)}_{a,b,0} = \frac{(2 a+3 b) (a+1)_{b-1}}{2 \Gamma (b+1)}\,,\qquad
c^{(0)}_{a,b,m} = - \sum\limits_{k=0}^{\lfloor\frac{b-1}{2}\rfloor} \frac{\Gamma(3k+1)}{\Gamma(k+1)\Gamma(2k+2)}c^{(0)}_{a+1+3k,b-1-2k,m-1} \,,
\label{c}
\end{equation}
which we can use to write an expression for $c^{(0)}_{a,b,m}$ in terms of $c^{(0)}_{a,b,0}$
\begin{equation}
c^{(0)}_{a,b,m} = (-1)^m \sum\limits_{k_1,\ldots,k_m=0}^\infty
c^{(0)}_{a+m+3k,b-m-2k,0}
\prod\limits_{i=1}^m \frac{\Gamma(3k_i+1)}{\Gamma(k_i+1)\Gamma(2k_i+2)}\,, \quad k=k_1+\ldots+k_m\,.
\label{cm_to_c0}
\end{equation}
Next we sum the terms above over $a$ and $b$ for several fixed values of $k$ and $m$ and guess the general form
\begin{equation}
\sum\limits_{a,b=0}^\infty c^{(0)}_{a+m+3k,b-m-2k,0} x^a y^b =
\frac12 \frac{y+2}{1-x-y} \frac{y^{2k+m}}{(1-y)^{3k+m}}\,.
\end{equation}
Inserting this into \eqref{cm_to_c0}, the sums over $k_i$ factorise and we find
\beq
\sum\limits_{a,b=0}^\infty c^{(0)}_{a,b,m} x^a y^b =
\frac12 \frac{y+2}{1-x-y} \left(\frac{\sqrt{1-4y}-1}{2} \right)^m\,,
\eeq
a formula that was already found in \cite{Zagier:2019eus}.
In order to perform the sum over $m$, we consider the generating series found in \cite{Green:2008uj,Zagier:2019eus}
\beq
\sum\limits_{m=0}^{\infty} F^{(0)}_m(\delta) \left(\frac{z}{\delta} \right)^m =  \binom{z+\delta-1}{\delta-1}^2\,.
\eeq
Combining everything, we can compute the sum
\bea
2\sum\limits_{a,b=0}^\infty \hat\sigma_2^a \hat\sigma_3^b \alpha^{(0)}_{a,b} ={}&
\sum\limits_{a,b,m=0}^\infty \sum\limits_{\delta=1}^\infty \frac{2}{ \delta^3} x^a y^b c^{(0)}_{a,b,m}  F^{(0)}_{m} (\delta)\\
={}& \sum\limits_{m=0}^\infty \sum\limits_{\delta=1}^\infty \frac{1}{\delta^3} \frac{y+2}{1-x-y} \left(\frac{\sqrt{1-4y}-1}{2} \right)^m  F^{(0)}_{m} (\delta)\\
={}& \sum\limits_{\delta=1}^\infty \frac{1}{\delta^3} \frac{y+2}{1-x-y} \binom{z+\delta-1}{\delta-1}^2\,,
\eea{summing_alpha}
with
\beq
x = \frac{\hat\sigma_2}{\delta^2}\,, \qquad
y = \frac{\hat\sigma_3}{\delta^3}\,, \qquad
z = \frac{\delta}{2} \left( \sqrt{1-4y}-1\right)\,.
\label{xyz}
\eeq
One can now use the formula
\beq
\binom{a-1}{b-1}^2 = \prod\limits_{j=1}^{b-1} \left(1- \frac{a(b-a)}{j(b-j)} \right)\,,
\label{binomial_squared}
\eeq
and insert the definition of $\hat\sigma_2$ and $\hat\sigma_3$
to find
\beq
A^{(0)}(S,T)
=\frac{1}{STU} +  \sum\limits_{\delta=1}^\infty \frac{1}{\delta^3}
\left( \frac{S}{\delta-S}+ \frac{T}{\delta-T}+ \frac{U}{\delta-U} + 2 \right)
\prod\limits_{j=1}^{\delta-1} \left( 1 - \frac{S T U}{j(\delta-j)\delta} \right)\,.
\label{alphasum_final}
\eeq
A proof that this matches the familiar result \eqref{flatspace_amplitude}
\beq
A^{(0)}(S,T) = -\frac{ \Gamma \left(- S\right) \Gamma \left(-T\right) \Gamma \left(- U \right) }{\Gamma \left(S +1\right) \Gamma \left( T+1\right) \Gamma \left( U +1\right) }\,,
\eeq
was given in \cite{Zagier:2019eus}.

In order to do the analogous sum for $A^{(1)}(S,T)$ we need to determine the generating series for $T^{(2)}_m(\delta)$ and $F^{(2)}_m(\delta)$ which appear in the expression for $\alpha^{(1)}_{a,b}$ in \eqref{beta}.
By studying them for fixed values of $m$ and $\delta$ we noticed that they can be written in the form
\bea
\sum\limits_{m=0}^{\infty} T^{(2)}_m(\delta) \left(\frac{z}{\delta} \right)^m ={}&  \sum\limits_{n=0}^{\delta-1} g_n(\delta) \binom{z+\delta-\frac{n}2-1}{\delta-n-1}^2\,,\\
\sum\limits_{m=0}^{\infty} F^{(2)}_m(\delta) \left(\frac{z}{\delta} \right)^m ={}&  \sum\limits_{n=0}^{\delta-1} h_n(\delta) 
\binom{z+\delta-\frac{n}2-1}{\delta-n-1}^2\,,
\eea{TF2_generating_series}
with
\bea
g_n(\delta) ={}& \frac{\delta}{(\delta-n)_n} \tilde g_n(\delta)\,, \qquad
h_n(\delta) = \frac{\delta}{(\delta-n)_n} \tilde h_n(\delta)\,,\\
\tilde g_0(\delta) ={}& \frac{1}{2 \delta} \left(3 \delta ^2-\delta +2\right)\,,\qquad
\tilde g_1(\delta) = \frac{1}{72} \left(\delta ^2+4 \delta +6\right)\,,\\
\tilde g_2(\delta) ={}& \frac{1}{7200}\left(-14 \delta ^3+3 \delta ^2+222 \delta +210\right)\,,\\
\tilde h_0(\delta) ={}& \frac{1}{96 \delta} \left(-112 \delta ^3+1872 \delta ^2-2216 \delta +1671\right)+2 \delta^2 \zeta(3) \,,\\
\tilde h_1(\delta) ={}&  \frac{1}{432} \left(-61 \delta ^3-703 \delta ^2-1221 \delta +2934\right) \,,\\
\tilde h_2(\delta) ={}&  \frac{1}{216000}\left(4598 \delta ^4+33699 \delta ^3-103219 \delta ^2-297105 \delta +581850\right)\,,
\eea{gh_examples}
and so on. We include further cases in a Mathematica notebook.
Now the sums can be done analogously to \eqref{summing_alpha}, with polynomials in $a,b,m$ turning into differential operators acting on $x^a$, $y^b$ or $z^m$.
In this way we obtain
\begin{align}
{}&A^{(1)}(S,T) + \frac{2}{3} \frac{\hat\sigma_2}{\hat\sigma_3^2}
= 2 \sum\limits_{a,b=0}^\infty \hat\sigma_2^a \hat\sigma_3^b    \alpha^{(1)}_{a,b}\nonumber\\
={}&\sum\limits_{a,b,m=0}^\infty \sum\limits_{\delta=1}^\infty \frac{2x^a y^b}{ \delta^4} \left( c^{(0)}_{a,b,m} \left(F^{(2)}_{m} (\delta) - (3+2a+3b)T^{(2)}_{m} (\delta)  \right)  
+ c^{(2,0)}_{a,b,m} F^{(0)}_m(\delta)+ c^{(2,1)}_{a,b,m} F^{(0)}_{m+1}(\delta)\right)\nonumber\\
={}& \sum\limits_{\delta=1}^\infty \sum\limits_{n=0}^{\delta-1} \frac{1}{ \delta^4} \mathcal{D}_n(\delta) \frac{y+2}{1-x-y} 
\binom{z+\delta-\frac{n}2-1}{\delta-n-1}^2\,,\label{summing_beta}
\end{align}
with the differential operator $\mathcal{D}_n(\delta)$ given by (note that $\partial_z$ only acts on $z$ but $\partial_y$ acts on $y$ and on $z$ through its definition)
\begin{align}
{}&\mathcal{D}_n(\delta) = h_n (\delta) - g_n (\delta) \left( 3 + 2 x \partial_x + 3 y \partial_y \right)+\delta_{n,0} \big( -\left(\tfrac{1}{2}x \partial_x+\tfrac{3}{4} y \partial_y+\tfrac{27}{4}\right) (z+\delta) \partial_z z \partial_z\nonumber \\
&-6 (x \partial_x)^2 y \partial_y-9 x \partial_x (y \partial_y)^2-16 x \partial_x y \partial_y-\tfrac{4}{3} (x \partial_x)^3-\tfrac{16}{3} (x \partial_x)^2-\tfrac{1}{3}x \partial_x-\tfrac{9}{2} (y \partial_y)^3\nonumber\\
&-12 (y \partial_y)^2-\tfrac{1}{2}y \partial_y-\tfrac{277}{32}
+ \left(6 x \partial_x y \partial_y+2 (x \partial_x)^2+\tfrac{26}{3} x \partial_x+\tfrac{9}{2} (y \partial_y)^2+13 y \partial_y-\tfrac{21}{8}\right)z \partial_z\nonumber\\
&+\left(3 x \partial_x y \partial_y+(x \partial_x)^2+\tfrac{55}{12} x \partial_x+\tfrac{9}{4} (y \partial_y)^2+\tfrac{55}{8} y \partial_y+\tfrac{33}{16}\right) \delta \partial_z \big)\,.
\label{D_operator}
\end{align}
Equation \eqref{summing_beta} has several nice properties. One can use \eqref{binomial_squared} to check that
\beq
\binom{z+\delta-\frac{n}2-1}{\delta-n-1}^2 
=\prod\limits_{j=1}^{\delta-n-1} \left( 1 - \frac{n^2-2n \delta+4\delta^2 y}{4j(\delta-n-j)} \right)\,,
\eeq
is a polynomial in $y$.
By expanding around the location of the poles we can also find general expressions for the residues
\beq
A^{(1)}(S,T) + \frac{2}{3} \frac{\hat\sigma_2}{\hat\sigma_3^2}
= \sum\limits_{k=1}^4 \frac{R_k(T,\delta)}{(S-\delta)^k} + O((S-\delta)^0)\,, \qquad \delta=1,2,\ldots\,.
\label{A1_residues}
\eeq
In terms of the function
\beq
R(T,\delta) = -\frac{\Gamma (T+\delta )^2}{\Gamma (\delta )^2 \Gamma (T+1)^2}\,,
\eeq
we find that the residues are the following polynomials in $T$
\begin{align}
R_4(T,\delta) ={}& R(T,\delta)\,,\nonumber\\
R_3(T,\delta) ={}& \frac{1}{6\delta} \left( 3 \delta  \partial_T-4 \right) R(T,\delta) \,,\nonumber\\
R_2(T,\delta) ={}& \frac{1}{\delta^2} \sum\limits_{n=0}^{\delta-1} g_n(\delta)  R(T+\tfrac{n}{2},\delta-n)
-\frac{
3 T (\delta +T) \partial_T^2+\tfrac12 (31 \delta +46 T) \partial_T +32}{12\delta^2} 
 R(T,\delta)  \,,\nonumber\\
R_1(T,\delta) ={}& \frac{1}{\delta^3} \sum\limits_{n=0}^{\delta-1}
\left( h_n(\delta) - g_n(\delta) \delta^2 (2+T\partial_T) \right)  R(T+\tfrac{n}{2},\delta-n) \label{residues}\\
&+ \frac{1}{96\delta^3} \left(
8 T^3 \partial_T^3 -4T(119 \delta +90 T) \partial_T^2-2(311 \delta +402 T)  \partial_T-927 
\right) R(T,\delta)\,, \nonumber
\end{align}
where
\bea
\partial_T R(T,\delta) ={}& 2 G_1(T,\delta ) R(T,\delta) \,, \\
\partial_T^2 R(T,\delta) ={}& -2 ( G_2(T,\delta ) - 2 G_1(T,\delta ){}^2) R(T,\delta) \,,\\
\partial_T^3 R(T,\delta) ={}& 4 ( G_3(T,\delta )-3 G_2(T,\delta ) G_1(T,\delta ) +2 G_1(T,\delta ){}^3) R(T,\delta) \,,\\
G_k(T,\delta ) ={}& H^{(k)}(T+\delta-1) - H^{(k)}(T) = \sum\limits_{j=1}^{\delta-1} \frac{1}{(T+j)^k}\,,
\eea{res_derivatives}
is given in terms of generalised harmonic numbers.

From \eqref{A1_residues} we see that $F^{(2)}_m(\delta)$ contributes only to single poles and
$T^{(2)}_m(\delta)$ contributes also to double poles.
The remaining terms have poles up to fourth order and the whole expression \eqref{summing_beta} has no simultaneous poles in different Mandelstams.
The pole of fourth order might be surprising if one expects the poles to arise from expanding a single pole in $1/\lambda$.
We show in appendix \ref{app:leading_poles} that to any order in $1/\lambda$, the pole of the highest order arises purely from the dispersive sum rule and does not depend on corrections to the OPE data. We also resum the  in $1/\lambda$ expansion for these poles.

We could now apply the inverse of the flat space transform \eqref{flat}
\beq
M(s_1,s_2)=
\text{FS}^{-1}(A(S,T)) = \frac{1}{2\lambda^\frac{3}{2} c}
\int\limits_{0}^\infty d\beta e^{-\beta} \beta^5
A\left( \frac{\beta s_1}{2\sqrt{\lambda}} ,\frac{\beta s_2}{2\sqrt{\lambda}}  \right)\,,
\eeq
to obtain the summed Mellin amplitude. As discussed in \cite{Penedones:2010ue}, the poles of $A(S,T)$ will lead to exponential integrals and hence branch cuts. These originate from many poles of the non-perturbative Mellin amplitude which have vanishing separation at large $\lambda$, but become separated when applying the flat space transform.

\section{Conclusions}
\label{sec:conclusions}

In this paper we determined the full $1/\sqrt{\lambda}$ contribution to the Virasoro-Shapiro amplitude on AdS$_5 \times$ S$^5$,
by solving the dispersive sum rules derived in \cite{Alday:2022uxp},
using the crucial assumption that the Wilson coefficients are single-valued periods. The resulting correction possesses an analytic structure which naturally generalises that of the Virasoro-Shapiro amplitude in flat space. 

The natural next step is to determine the next layer of Wilson coefficients $\alpha^{(2)}_{a,b}$ with similar arguments. A preliminary study shows that single-valuedness is also powerful enough to determine the coefficients $\alpha^{(2)}_{a,b}$ uniquely, once quantities like $\langle f_0(\de,\ell) \tau_2(\de,\ell)^2 \rangle$ are provided. This would require solving a mixing problem to order $1/\sqrt{\lambda}$, considering more general correlators. Single-valuedness is not powerful enough if we treat these quantities as unknown, not surprisingly. 

Certain universal parts of the answer, which do not depend on corrections to the CFT data, can be studied to all orders in $1/\lambda$, see appendix \ref{app:leading_poles}. It would be interesting to explore this further.

Over the last few years there has been great progress in understanding how supersymmetry, via localisation results, gives integrated constraints for the correlator under consideration \cite{Binder:2019jwn,Chester:2020dja,Chester:2019jas,Chester:2020vyz,Dorigoni:2021bvj,Dorigoni:2021guq,Collier:2022emf}. In the present context, this will lead to two linear constraints for the Wilson coefficients at each order in $1/\lambda$. These linear constrains are written in terms of single zeta values of odd arguments, and it would be interesting to understand how this arises from our procedure.

Combining these two linear constraints and the flat space limit with our new solution for $\alpha^{(1)}_{0,1}$, we can for the first time fully determine the $D^8 R^4$ term at planar order, which appears at $O(1/\lambda^{\frac72})$ and depends on four Wilson coefficients
\begin{align}
{}&M(s_1,s_2) = \frac{8}{(s_1-\frac23) (s_2-\frac23) (s_3-\frac23)}
+ \frac{120 \zeta (3)}{\lambda ^{3/2}}
+ \frac{210 \left(3 \sigma _2+7\right) \zeta (5)}{\lambda ^{5/2}}+ \label{M_with_D8R4}\\
&\frac{140 \left(108 \sigma_3 - 99 \sigma _2-320\right) \zeta (3)^2}{3 \lambda ^3}
+ \frac{35 \left(2592 \sigma _2^2-77328 \sigma _3+73638 \sigma _2+178909\right) \zeta (7)}{16 \lambda ^{7/2}}
+O(\lambda^{-4})\,.\nonumber
\end{align}

Among the CFT data provided by our solution we reproduce the anomalous dimensions of short Konishi-like operators, in full agreement with the results from integrability, together with their structure constants. It would be interesting to reproduce these structure constants from integrability methods, along the lines of \cite{Basso:2022nny}. A related direction is the interplay between integrability and the conformal bootstrap, explored first in \cite{Cavaglia:2021bnz,Cavaglia:2022qpg} and in our context in \cite{Caron-Huot:2022sdy}. Now that our analytic methods allow to explore $1/\lambda$ corrections, it would be interesting to feed this into the program of \cite{Caron-Huot:2022sdy}.

Finally, we hope that our results can fuel progress towards determining the worldsheet theory for strings on AdS$_5 \times$ S$^5$.
Recent progress on determining the vertex operators has been made in \cite{Fleury:2021ieo}.
It would be very interesting to see explicitly how the expression for the AdS Virasoro-Shapiro amplitude, in a $1/\lambda$ expansion, arises from the worldsheet theory.

\section*{Acknowledgements} 

We thank Michael Green for useful discussions.
The work of LFA and TH is supported by the European Research Council (ERC) under the European Union's Horizon
2020 research and innovation programme (grant agreement No 787185). LFA is also supported in part by the STFC grant ST/T000864/1.
The work of JS is supported by an early postdoc mobility fellowship (grant P2ELP2199748) by the Swiss National Science Foundation (SNSF).
The authors would like to acknowledge the use of the University of Oxford Advanced Research Computing (ARC) facility in carrying out this work. http://dx.doi.org/10.5281/zenodo.22558

\appendix

\section{Crossing-symmetric dispersion relations}
\label{app:crossing-symmetric_dispersion_relations}

Based on earlier work in \cite{Auberson:1972prg,Sinha:2020win}, a crossing symmetric dispersion relation for Mellin amplitudes was derived in \cite{Gopakumar:2021dvg}. The idea is to use the variables
\beq
a = \frac{y}{x} \,, \qquad
\hat s_k(z,a) = a- \frac{a(z-z_k)^3}{z^3-1}\,, \qquad z_k = e^{\frac23 \pi i (k-1)}\,, \quad k=1,2,3\,, 
\label{az}
\eeq
and in this appendix we used the rescaled variables
\beq
\hat s_k = \frac{s_k}{2}\,, \qquad
x = \frac{\sigma_2}{8}\,, \qquad
y= - \frac{\sigma_3}{8}\,.
\eeq
One can derive a fixed-$a$ dispersion relation by deforming an integration contour in $z$.
The derivation of the dispersion relation is described in detail in \cite{Auberson:1972prg} and in the supplementary material of \cite{Sinha:2020win}.
One starts with the expression
\beq
M(z,a) = \frac{1}{2\pi i} \frac{z^3}{1-z^3} \oint_z dz' \frac{z'^3-1}{z'^3(z'-z)} M(z',a) + \text{const.}\,,
\label{dispersion1_start}
\eeq
and deforms the integration contour, picking up the poles corresponding to OPE singularities, which lie on the unit circle (at least if $a$ lies in a certain range).
A constant contribution is not determined by this relation because of poles at $z=0$ and $z=\infty$. The Regge limit $|s_k| \to \infty$ is mapped to the three roots of unity $z=z_k$, and one assumes the following bound in the Regge limit
\beq
M(s_1,s_2) = o(s_1^2) \ \text{for} \ |s_1|\to \infty\,, \quad s_2 \ \text{fixed}\,,
\eeq
which amounts to
\beq
M(z,a) = o\left(\frac{1}{(z-z_k)^2}\right) \ \text{for} \ |z|\to z_k\,, \quad a \ \text{fixed}\,.
\eeq
The factor $(z'^3-1)/z'^3$ in \eqref{dispersion1_start} is a subtraction that ensures that these singularities do not contribute.
In this way one finds the following crossing symmetric expression for the Mellin amplitude in terms of OPE data and Mack polynomials
\beq
M(s_1,s_2) = \sum\limits_{\tau,\ell,m} \frac{c_{\tau,\ell}^{(m)}}{\tau_m} P_{\tau,\ell}(\tau_m,\hat s_2'(\tau_m,a))  H(\tau_m,\hat s_1,\hat s_2,\hat s_3) + \text{ const.}\,,
\label{dispersion1}
\eeq
where $\tau_m = \tau/2+m-2/3$, 
\bea
c^{(m)}_{\tau,\ell} ={}& 
=  \frac{2^{3\ell-7} \tau ^6 \Gamma \left(\ell+\frac{\tau }{2}+\frac{3}{2}\right) \Gamma \left(\ell+\frac{\tau }{2}+\frac{5}{2}\right) \Gamma \left(m+\frac{\tau }{2}-1\right)^2}{(\ell+1)
   \Gamma (m+1) \Gamma \left(\ell+\frac{\tau }{2}+2\right)^2 \Gamma (\ell+m+\tau +3)} f(\tau,\ell)\,,
\eea{ope_conventions}
is related to the OPE coefficient $f(\tau,\ell)$ from \eqref{OPEStringy2}
and
\bea
H(\tau_m,\hat s_1,\hat s_2,\hat s_3) &= \frac{\hat s_1}{\tau_m-\hat s_1}+\frac{\hat s_2}{\tau_m-\hat s_2}+\frac{\hat s_3}{\tau_m-\hat s_3}\,,\\
\hat s_{2}^{\prime}\left(\tau_m, a\right)&=-\frac{\tau_m}{2}\left(1 - \left(\frac{\tau_m+3 a}{\tau_m-a}\right)^{1 / 2}\right)\,,\\
P_{\tau,\ell}(\tau_m,\hat s_3) &= 4^{-\ell} Q^{\tau+4,d}_{\ell,m} \left(2(\hat s_3+\tfrac23)-4\right)\,,
\eea{H_sprime}
where the Mack polynomial $Q_{\ell, m}^{\tau, d}(t)$ is defined as in \cite{Alday:2022uxp}.

\subsection{Wilson coefficients}
\label{app:wilson_coefficients}

An expression for the Wilson coefficients in the expansion of the Mellin amplitude
\beq
M(s_1,s_2) = \sum\limits_{a,b=0}^{\infty} = \cM_{a,b} x^a y^b\,,
\label{xy_expansion}
\eeq
in terms of OPE data
was derived in \cite{Gopakumar:2021dvg} by expanding
\eqref{dispersion1} in $x$ and $y$
\beq \label{Mdef}
\mathcal{M}_{a-b, b} = \sum_{\substack{\tau,\ell,m }}^{\infty}c_{\tau,\ell}^{(m)}\mathcal{B}_{a, b}^{(\tau,\ell,m)}\,, a \geq 1\,.
\eeq
Here $\mathcal{B}_{a, b}^{(\tau,\ell,m)}$ is given by
\beq\label{eq:BellnmU}
\mathcal{B}_{a, b}^{(\tau,\ell,m)}=\sum_{q=0}^{b}\mathfrak{U}^{\left(\tau_m\right)}_{a,b,q}(-1)^{b+q} P_{\tau, \ell;q}\left(\tau_m,0\right)\,,
\eeq
where $P_{\tau, \ell;q}\left(\tau_m,0\right)=\partial_{\hat s_2}^q P_{\tau, \ell}\left(\tau_m,\hat s_2\right)|_{\hat s_2=0}$ and 
\beq
\begin{split}
&\mathfrak{U}^{\left(\tau_m\right)}_{a,b,q}=\frac{(a-q-1)! (b+2 a-3q)}{q! (a-b)! (b-q)! \left(\tau_m\right)^{b+2 a-q+1}}\,_{4} F_{3}\left[\begin{array}{c}
\frac{q}{2}+\frac{1}{2},\frac{q}{2},q-b,q+1-\frac{2 a+b}{3} \\
q+1,q+1-a,q-\frac{2 a+b}{3}
\end{array} ; 4\right]\,.
\end{split}
\eeq

By equating $M(s_1,s_2)$ to our 
reduced Mellin amplitude \eqref{M} (minus the supergravity amplitude), the expansion
\beq
\cM_{a,b} = \sum\limits_{c=0}^{\infty} \frac{\cM_{a,b}^{(c)}}{\lambda^{\frac12(3+2a+3b+c/2)}} \,,
\label{Mab_expansion}
\eeq
is related to our conventions by
\beq
\alpha^{(k)}_{a,b} = \frac{(-1)^b \cM_{a,b}^{(2k)}}{\Gamma(6+2a+3b)}\,.
\label{alpha_of_M}
\eeq
Now we can insert the OPE data expansions \eqref{twistsStringy} and \eqref{OPEStringy2},
expand in large $\lambda$ and sum over $m$ (as discussed in \cite{Alday:2022uxp}) to obtain the expressions for $\alpha^{(0)}_{a,b}$ and $\alpha^{(1)}_{a,b}$ in \eqref{alpha} and \eqref{beta}, generalising the expressions in \cite{Alday:2022uxp}.

\subsection{Recursion relations}

For the reduced Mellin amplitude at hand, we can derive stronger crossing-symmetric dispersion relations and indeed recursion relations for the Wilson coefficients, because the reduced Mellin amplitude satisfies the stronger Regge bound (we assume that for string theory the chaos bound is not saturated)
\beq
M(s_1,s_2) = o(s_1^{-2}) \ \text{for} \ |s_1|\to \infty\,, \quad s_2 \ \text{fixed}\,,
\eeq
or in terms of $z$ and $a$
\beq
M(z,a) = o\left((z-z_k)^{2}\right) \ \text{for} \ |z|\to z_k\,, \quad a \ \text{fixed}\,.
\eeq
This means we do not need the subtraction used above, and can even make use of the fact that the Mellin amplitude vanishes at $z=z_k$.
The trick to get a nice relation is to use the combination
\beq
\frac{z^3}{(z^3-1)^2} = - \frac{x}{27a^2}\,.
\eeq
We first consider the expression
\beq
M(z,a) = \frac{1}{2\pi i} \frac{(z^3-1)^2}{z^3}\frac{z^3}{1-z^3} \oint_z dz' \frac{z'^3}{(z'^3-1)^2} \frac{z'^3-1}{z'^3(z'-z)} M(z',a)\,,
\label{dispersion2_start}
\eeq
which results in the following dispersion relation, that now also fixes the constant part of the Mellin amplitude (that means \eqref{Mdef} is also valid for $a=0$)
\beq
M(s_1,s_2) = \sum\limits_{\tau,\ell,m} \frac{c_{\tau,\ell}^{(m)}}{\tau_m} P_{\tau,\ell}(\tau_m,\hat s_2'(\tau_m,a)) \frac{\tau_m^3}{x(\tau_m-a)} H(\tau_m,\hat s_1,\hat s_2,\hat s_3) \,.
\label{dispersion2}
\eeq
The Regge bound is even strong enough to allow another 'addition' so we can start with
\beq
M(z,a) = \frac{1}{2\pi i} \frac{(z^3-1)^4}{z^6}  \frac{z^3}{1-z^3} \oint_z dz' \frac{z'^6}{(z'^3-1)^4}\frac{z'^3-1}{z'^3(z'-z)} M(z',a)\,,
\label{dispersion3_start}
\eeq
and find the relation
\beq
M(s_1,s_2) = \sum\limits_{\tau,\ell,m} \frac{c_{\tau,\ell}^{(m)}}{\tau_m} P_{\tau,\ell}(\tau_m,\hat s_2'(\tau_m,a)) \left(\frac{\tau_m^3}{x(\tau_m-a)}\right)^2 H(\tau_m,\hat s_1,\hat s_2,\hat s_3) \,.
\label{dispersion3}
\eeq

We can use the fact that we have two dispersion relations to obtain recursion relations. To this end we can combine the expansion in $x$ and $y$ with the one in OPE data and write
\beq
\cM_{a,b} = \sum\limits_{\tau,\ell,m} \cM_{a,b}(\tau,\ell,m)\,.
\eeq
Equating \eqref{dispersion2} and  \eqref{dispersion3} we find
\beq
M(s_1,s_2) = \sum_{a,b}\sum\limits_{\tau,\ell,m} \cM_{a,b}(\tau,\ell,m) x^a y^b
= \sum_{a,b}\sum\limits_{\tau,\ell,m} \frac{\tau_m x - y}{\tau_m^3} \cM_{a,b}(\tau,\ell,m) x^a y^b\,,
\eeq
which implies the recursion relation
\beq
\sum\limits_{\tau,\ell,m}\cM_{a,b}(\tau,\ell,m) = \sum\limits_{\tau,\ell,m} \frac{1}{\tau_m^2} \cM_{a-1,b}(\tau,\ell,m)- \frac{1}{\tau_m^3} \cM_{a,b-1}(\tau,\ell,m)\,.
\label{M_recursion}
\eeq
The next step is to expand in $1/\lambda$ and do the sums over $m$. To leading order this implies the recursion relation
\beq
\alpha^{(0)}_{a,b}(\delta) = \frac{1}{\delta^2} \alpha^{(0)}_{a-1,b}(\delta) +  \frac{1}{\delta^3} \alpha^{(0)}_{a,b-1}(\delta)\,,
\label{alpha_recursion}
\eeq
where $\alpha^{(0)}_{a,b}(\delta)$ is the summand of $\alpha^{(0)}_{a,b}$
\beq
\alpha^{(0)}_{a,b} = \sum\limits_{\delta=1}^\infty \alpha^{(0)}_{a,b}(\delta)\,.
\eeq
One checks that \eqref{alpha} is a solution of this recursion for the boundary condition
$\alpha^{(0)}_{-1,b}(\delta) = 0$. As the boundary data $\alpha^{(0)}_{-b,b}(\delta)$ is non-trivial ($\alpha^{(0)}_{-b,b}$ vanishes only after summing over $\delta$), the recursion can be seen as a neat way to encode the $a$ dependence of \eqref{alpha}.

\section{Alternative representation for spin sums}
\label{app:alternative_representation}

The formulas \eqref{TF2_generating_series} imply that there is an alternative representation for $T^{(2)}_m(\delta)$ and $F^{(2)}_m(\delta)$. Adapting a computation from \cite{Green:2008uj}
\begin{align}
\sum\limits_{m=0}^{\infty} T^{(2)}_m(\delta) \left(\frac{z}{\delta} \right)^m ={}&  \sum\limits_{n=0}^{\delta-1} g_n(\delta) \binom{z+\delta-\frac{n}2-1}{\delta-n-1}^2
\nonumber\\
={}&  \sum\limits_{n=0}^{\delta-1} g_n(\delta) \prod\limits_{k=1}^{\delta-n-1} \left( 1 + \frac{z+\frac{n}{2}}{k}\right)^2
\nonumber\\
={}&  \sum\limits_{n=0}^{\delta-1} g_n(\delta) \prod\limits_{k=1}^{\delta-n-1} \left( 1 + 4 \sum\limits_{s \in \{1,2\}} \left(\frac{z+\frac{n}{2}}{2k}\right)^s \right)
\nonumber\\
={}&  \sum\limits_{n=0}^{\delta-1} g_n(\delta) \sum\limits_{d=0}^\infty
\sum\limits_{\substack{\delta-n>k_1>\ldots> k_d > 0\\s_1,\ldots,s_d \in \{1,2\}}}
\frac{2^{2d-s_1-\ldots-s_d} (z+\frac{n}{2})^{s_1+\ldots+s_d} }{k_1^{s_1} \ldots k_d^{s_d}}
\label{T_rep_from_generating_series}\\
={}&  \sum\limits_{n=0}^{\delta-1} g_n(\delta) \sum\limits_{d=0}^\infty
\sum\limits_{s_1,\ldots,s_d \in \{1,2\}}
2^{2d-s_1-\ldots-s_d} \left(z+\frac{n}{2}\right)^{s_1+\ldots+s_d} Z_{s_1,\ldots,s_d}(\delta-n-1) 
\nonumber\\
={}&  \sum\limits_{n=0}^{\delta-1} g_n(\delta) \sum\limits_{w=0}^\infty
\left(z+\frac{n}{2}\right)^{w} \frac{F^{(0)}_w (\delta-n)}{(\delta-n)^{w}}
\nonumber\\
={}&  \sum\limits_{n=0}^{\delta-1} g_n(\delta) \sum\limits_{w=0}^\infty
\sum\limits_{m=0}^w
\binom{w}{m} \left(\frac{n}{2}\right)^{w-m}  z^m \frac{F^{(0)}_w (\delta-n)}{(\delta-n)^{w}}\,,
\nonumber
\end{align}
we can read off
\beq
T^{(2)}_m(\delta) = \delta^m
\sum\limits_{n=0}^{\delta-1} g_n(\delta) \sum\limits_{w=m}^{2(\delta-n-1)}
\binom{w}{m} \left(\frac{n}{2}\right)^{w-m}   \frac{F^{(0)}_w (\delta-n)}{(\delta-n)^{w}}\,,
\eeq
and similar for $F^{(2)}_m(\delta)$. We used in the final formula that $F^{(0)}_w (\delta-n)$ vanishes for $w > 2(\delta-n-1)$.

\section{More Bootstrap Constraints}\label{app:More Bootstrap Constraints}

We have explored two types of constraints that the bootstrap imposes on heavy operators dual to short strings. One constraint is the fact that the coefficients in the low energy expansion of the Mellin amplitude are related to dimensions and OPE coefficients of heavy operators. Schematically,
\begin{align}\label{Generic xi}
\xi_{a, b} = \sum_{\delta=1}^{\infty} \frac{p_{\xi, b}(a, \delta)}{\delta^{r+2a}},
\end{align}
where $\xi_{a, b}$ is some generic coefficient in the low energy expansion of the Mellin amplitude and the function $p_{\xi, b}(a, \delta)$ depends on the OPE data through a sum over spins. It is expected that $\xi_{a, b}$ is a single valued period, and so this constrains the CFT data.
Another constraint that the bootstrap imposes on the heavy operators is the equation
\begin{align}
\xi_{a, b} =0, ~~~ a=-b, ~... ~, -1.
\end{align}

In this section, we explore a third type of constraint. The essential idea is that, at strong coupling, the Mellin amplitude should be similar to a tree level string amplitude. Tree level string amplitudes obey stringent bootstrap constraints, see \cite{Caron-Huot:2016icg, Sever:2017ylk, Figueroa:2022onw, Geiser:2022icl, Coon:1969yw}. 
We take advantage of this in the following way. We write down a two sided dispersion relation, using the $s_1$ and $s_2$ channels. This dispersion relation stops converging for values of $\Re(s_3)$ sufficiently high, i.e.\ when we reach the first pole in $s_3$. Furthermore, the two sided dispersion relation must diverge in a precise way, such that it reproduces the residue in $s_3$. This gives nontrivial constraints.

\subsection{Flat Space}

This idea is easier to express in flat space. The Virasoro-Shapiro amplitude is given by
\begin{align}
A^{(0)}(S, T) = -\frac{ \Gamma \left(- S\right) \Gamma \left(- T\right) \Gamma \left(S+T \right) }{\Gamma \left( 1+S\right) \Gamma \left( 1+T\right) \Gamma \left(1-S - T\right) }\,.\label{flatspace_amplitude_app}
\end{align}
We ask the following question. Without knowing the form of the amplitude (\ref{flatspace_amplitude_app}), and assuming only the spectrum of particles exchanged (mass and spin) and Regge boundedness, i.e.\ 
\begin{align}\label{FlatSpaceRegge}
\lim_{S\rightarrow i \infty} |A^{(0)}(S, T)| \leq \frac{1}{|S|^2}, ~~~ \Re(T)<0,
\end{align}
what can we say about particle couplings?

We write a two sided dispersion relation
\bea
A^{(0)}(S, T) ={}& -\frac{1}{U^2}\left(\frac{1}{S}+\frac{1}{T}\right) 
- \sum_{\delta = 1}^{\infty} \sum_{\ell=0,2}^{2 \delta -2} \frac{ f_0(\delta, \ell) C_{\ell}^{(1)}(1+ \frac{2U }{ \delta})}{(1 + \ell) \delta^2} \left(\frac{1}{S-  \delta} + \frac{1}{T-  \delta}\right) \, ,
\eea{VStwosided}
where $U=-S-T$. By parametrising in this manner, the particle couplings $f_0(\delta, \ell)$ turn out to be numerically equal to the OPE coefficients that enter in (\ref{OPEStringy2}). This is due to the flat space transform (\ref{flat}). Also, we simply wrote down the part of the amplitude 
\beq
\frac{1}{S T U} = -\frac{1}{U^2}\left(\frac{1}{S}+\frac{1}{T}\right)\,,
\eeq
because it won't play any role.
The dispersion relation converges for $\Re(U)<1$. For $\Re(U)=1$, it must diverge, in such a way as to give rise to the correct pole and residue in $U$. One very natural way this can happen is if
\begin{align}\label{Asymptoticsf0}
\lim_{\delta \rightarrow \infty}  \sum_{\ell=0,2}^{2 \delta -2}  \frac{ f_0(\delta, \ell) C_{\ell}^{(1)}(1+ \frac{2U }{ \delta})}{2(1 + \ell) \delta^3} =  f_0(\delta=1, \ell=0) \times \kappa \times \delta^{\kappa(U-1)-1}\,,
\end{align}
where $\kappa$ is some number that is not fixed by this reasoning. The reason for this asymptotic is that
\begin{align}
f_0(\delta=1, \ell=0) \times \kappa \times \int_{\delta_0}^{\infty} d \delta \delta^{\kappa (U-1)-1} \underbrace{=}_{U \rightarrow 1}- f_0(\delta=1, \ell=0) \frac{1}{U-1}.
\end{align}
This leads to the equation
\begin{align} \label{Asymptoticsf0 2}
\boxed{\lim_{\delta \rightarrow \infty} \sum_{\ell=0,2}^{2 \delta -2} \frac{f_0(\delta, \ell) C_{\ell}^{(1)}(1+ \frac{2 }{\delta})}{(1 + \ell) \delta^2} = \kappa\,,}
\end{align}
where $\kappa$ is some positive number, whose value is not fixed by this reasoning. When we plug the numerical values of $f_0(\delta, \ell)$, it turns out that (\ref{Asymptoticsf0 2}) is precisely obeyed for every $\delta \in \mathbb{N}$, for $\kappa =1$.

\subsection{Flat Space from AdS}

In this section we show how to derive (\ref{Asymptoticsf0 2}) from: Mellin amplitudes in AdS, CFT Regge boundedness and formulas (\ref{twistsStringy}) and (\ref{OPEStringy2}) for the spectrum of exchanged operators. Our discussion is similar to the one in \cite{Penedones:2010ue} (though not exactly the same). The point of going through this is to later use the same line of reasoning to derive a generalisation of (\ref{Asymptoticsf0 2}) for the case involving $f_2(\delta, \ell)$ and $\tau_2(\delta, \ell)$, see formula (\ref{HugeUgly}).
The starting point is the dispersion relation
\begin{align}\label{dispRel}
M(s_1, s_2) = \sum_{\text{stringy}} C_{\tau, \ell}^2 \omega_{\tau, \ell}(s_1, s_2),
\end{align}
where
\begin{align}
\omega_{\tau, \ell}(s_1, s_2) &\equiv \sum_{m=0}^{\infty} \mathcal{Q}_{\ell, m}^{\tau +4, d=4}\left(s_3 - \frac{8}{3}\right) \left( \frac{1}{s_1+ \frac{4}{3} -\tau - 2m} + \frac{1}{s_2+ \frac{4}{3}-\tau - 2m} \right), \\
\mathcal{Q}_{\ell, m}^{\tau, d}(t) &\equiv -\frac{2^{3 \ell+2 \tau -2} \Gamma \left(\ell+\frac{\tau }{2}-\frac{1}{2}\right) \Gamma \left(\ell+\frac{\tau }{2}+\frac{1}{2}\right) \Gamma \left(-\frac{d}{2}+\ell+\tau +1\right) Q_{\ell, m}^{\tau, d}(t)}{\pi  \Gamma (m+1) \Gamma \left(\ell+\frac{\tau }{2}\right)^2 \Gamma (\ell+\tau -1) \Gamma \left(-m-\frac{\tau }{2}+4\right)^2 \Gamma \left(-\frac{d}{2}+\ell+m+\tau +1\right)}.  \nonumber
\end{align}
$Q_{\ell, m}^{\tau, d}(t)$ is a Mack polynomial, see \cite{Alday:2022uxp} for our conventions.
This dispersion relation is valid at finite $\lambda$ and also when the Mellin variables are finite. We now take the limit $s_1 \rightarrow s_1 \lambda^{\frac{1}{2}}, s_2 \rightarrow s_2 \lambda^{\frac{1}{2}}$ and $\lambda \rightarrow\infty$. As explained in \cite{Penedones:2010ue}, this is what controls the flat space limit of AdS. In this regime, the sums in $m$ are dominated by terms of order $\tau^2 \sim \lambda^{\frac{1}{2}}$. We can thus do the substitution $m=4 \delta \lambda^{\frac{1}{2}} x$ and $\sum_{m=0}^{\infty} \rightarrow 4 \delta \lambda^{\frac{1}{2}} \int_0^{\infty} dx $ and we get
\begin{align}\label{rhsLargeLambda}
\lim_{\lambda \rightarrow \infty} M(\lambda^{\frac{1}{2}} s_1, \lambda^{\frac{1}{2}} s_2) &= \frac{1}{\lambda^{\frac{3}{2}}} M^{(0)}(s_1, s_2) + O\left(\frac{1}{\lambda^2}\right),  \\
M^{(0)}(s_1, s_2) &\equiv  \sum_{\delta=1}^{\infty} \sum_{\ell=0, 2}^{\infty} f_0(\delta , \ell) \int_0^{\infty} dx  \frac{e^{-\frac{1}{4 x}} \left(-\frac{1}{\frac{s_1}{4 \delta}-2 x}-\frac{1}{\frac{s_2}{4 \delta}-2 x}\right) C_{\ell}^{(1)}\left(\frac{s_3}{4 \delta x}+1\right)}{4096 (\ell+1)  \delta^3 x^6}\,.\nonumber
\end{align}
Note that Mack polynomials turn into Gegenbauer polynomials in this limit \cite{Costa:2012cb}. 

At strong coupling, the Mellin amplitude develops cuts, as can be seen by explicitly evaluating the above integrals for specific values of the spins. This is intuitive, because the location of the poles is at $s_1, s_2, s_3 = \tau + 2 m + \frac{4}{3}$, $m \in \mathbb{N}_0$, so if $\tau \sim \lambda^{\frac{1}{2}}$ is parametrically large, there is no way of distinguishing between consecutive poles and the sequence of poles basically condenses into a line.
Because of this, it is useful to introduce the transform
\begin{align}\label{FlatSpaceTransform}
A^{(0)}(S, T) =  2 \int_{\kappa - i \infty}^{\kappa + i \infty} \frac{d \alpha}{2 \pi i} \alpha^{-6} e^{\alpha} M^{(0)}\left(\frac{2S}{\alpha}, \frac{2T}{\alpha} \right).
\end{align}
This the flat space transform \eqref{flat} after having performed the large $\lambda$ expansion already in \eqref{rhsLargeLambda}. 
From (\ref{FlatSpaceTransform}) we conclude that $A^{(0)}(S, T)$ is bounded in the Regge limit like (\ref{FlatSpaceRegge}). It is crossing symmetric: $A^{(0)}(S, T)=A^{(0)}(T, S)=A^{(0)}(S, U)$.
Its poles are generated through the following mechanism. Let us focus on the poles in $S$. Insert (\ref{rhsLargeLambda}) into (\ref{FlatSpaceTransform}) and commute the $x$ and $\alpha$ integrals. We can deform the contour in $\alpha$ to the left due to the $e^{\alpha}$ in the integrand. The contour is bent in such a way that we avoid the $\alpha=0$ singularity, but we pick up the pole at $\alpha = \frac{S}{4 \delta x}$. The $x$ integral then becomes, up to a factor
\begin{align}\label{MechanismPole}
\int_0^{\infty} dx \frac{e^{-\frac{1}{4x} + \frac{S }{4 \delta x}}}{x^2} = -\frac{4 \delta}{S -\delta}\,.
\end{align}
We conclude that the poles and residues of $A^{(0)}(S, T)$ in $S$ are according to 
\begin{align}\label{Poles T0}
A^{(0)}(S, T) \approx -  \frac{ f_0(\delta, \ell) C_{\ell}^{(1)}(1+ \frac{2U }{\delta})}{(1 + \ell) \delta^2} \frac{1}{S- \delta} \,.
\end{align}
So, we have established from (\ref{FlatSpaceTransform}) the necessary ingredients to write a two sided dispersion relation for $T_0(S, T)$ and in this way obtain (\ref{Asymptoticsf0 2}).

\subsection{AdS constraints}

We can compute corrections to formula (\ref{rhsLargeLambda})
\begin{align}\label{rhsLargeLambdaMore}
\lim_{\lambda \rightarrow \infty} M(\lambda^{\frac{1}{2}} s_1, \lambda^{\frac{1}{2}} s_2) &= \frac{1}{\lambda^{\frac{3}{2}}} M^{(0)}(s_1, s_2) +  \frac{1}{\lambda^{2}} M^{(1)}(s_1, s_2) + \ldots
\end{align}	
The basic ingredient is to understand how to expand Mack polynomials. The formula we need is
\begin{align}\label{MackExpansion}
Q_{\ell, m=x \tau^2}^{\tau, d=4}\left(s_3 \lambda^{\frac{1}{2}} - \frac{8}{3} \right) = \lambda^{\frac{\ell}{2}} \left(\mathcal{C}(x, \delta, \ell; s_3) + \mathcal{C}_1(x, \delta, \ell; s_3) \frac{1}{\lambda^{\frac{1}{4}}} + \mathcal{C}_2(x, \delta, \ell; s_3) \frac{1}{\lambda^{\frac{1}{2}}} + ... \right),
\end{align}
where $\tau= 2 \sqrt{\delta} \lambda^{\frac{1}{4}} + \tau_1(\delta, \ell) + \tau_2(\delta, \ell)\lambda^{-\frac{1}{4}}+ \ldots$ and
\begin{align}
&\mathcal{C}(x, \delta, \ell; s_3) = 2^{\ell} x^{\ell} \delta^{\ell} C_{\ell}^{(1)}\left(1 + \frac{s_3}{4 x \delta}\right) , \\
&\mathcal{C}_1(x, \delta, \ell; s_3) =  \frac{1+4 \tau_1(\delta, \ell) x}{4 \sqrt{\delta}} \partial_x \mathcal{C}(x, \delta, \ell; s_3) \nonumber, \\
&\mathcal{C}_2(x, \delta, \ell; s_3) =  \frac{3 \ell+6 x^2+20 x}{48 \delta +96 \delta  \ell} \partial^2_x \mathcal{C}(x, \delta, \ell; s_3) + \frac{-12 x^2-40 x+3}{24 \ell+12} \partial_x \partial_{s_3} \mathcal{C}(x, \delta, \ell; s_3) \nonumber \\ 
& - \frac{1}{6 \delta} \partial_x \mathcal{C}(x, \delta, \ell; s_3)   +\frac{x \tau_2(\delta, \ell)}{\sqrt{\delta}} \partial_x \mathcal{C}(x, \delta, \ell; s_3) + \frac{x \tau^2_1(\delta, \ell)}{4 \delta} \partial_x \mathcal{C}(x, \delta, \ell; s_3) \nonumber \\ 
&+ \frac{\tau^2_1(\delta, \ell) x^2}{2 \delta} \partial^2_x \mathcal{C}(x, \delta, \ell; s_3) \nonumber + \frac{\tau_1(\delta, \ell)}{8 \delta} \partial_x  \mathcal{C}(x, \delta, \ell; s_3) + \frac{x \tau_1(\delta, \ell)}{4 \delta} \partial^2_x \mathcal{C}(x, \delta, \ell; s_3).  \nonumber
\end{align}
The flat space transform \eqref{flat} applied to \eqref{rhsLargeLambdaMore} implies
\begin{align}\label{FlatSpaceTransformS}
A^{(i)}(S, T) =  2 \int_{\kappa - i \infty}^{\kappa + i \infty} \frac{d \alpha}{2 \pi i} \alpha^{-6} e^{\alpha} M^{(i)}\left(\frac{2S}{\alpha} , \frac{2T}{\alpha} \right)\,.
\end{align}
In formula (\ref{rhsLargeLambdaMore}), we assumed that there was no correction proportional to $\frac{1}{\lambda^{\frac{7}{4}}}$. This is not automatic from  (\ref{dispRel}) and actually leads to the familiar conditions
\begin{align}
\tau_1(\delta, \ell) = - \ell-2\,, ~~ \< f_1(\delta, \ell) \> =  \<f_0(\delta, \ell)\> \frac{3 \ell + \frac{23}{4}}{\sqrt{\delta}}\,.
\end{align}
$A^{(1)}(S, T)$ inherits a two sided dispersion relation from $M^{(1)}(S, T)$
\begin{align}\label{TwoChannelT1}
A^{(1)}(S, T) = -\frac{2}{3} \frac{\hat\sigma_2}{\hat\sigma_3^2}+  \sum_{\delta =1}^{\infty}  \sum_{k=1}^{4} R_k(U,\delta) \left(\frac{1}{(S-  \delta)^k} + \frac{1}{(T-  \delta)^k}\right)\,, \qquad
-S-T < 1\,.
\end{align}
Note the appearance of poles of order $4$. The reason why we have higher poles is because the expansion in $\frac{1}{\sqrt{\lambda}}$ generates extra powers of $\frac{1}{x}$ which, upon doing an integral like (\ref{MechanismPole}), give rise to higher order poles. The residues involve $f_0(\delta, \ell)$, $f_2(\delta, \ell)$, $\tau_2(\delta, \ell)$, Gegenbauer polynomials and also derivatives of Gegenbauer polynomials.

\begin{align}
R_4(U,\delta) &= - \sum_{\ell=0, 2}^{2 \delta -2} \frac{f_0(\delta, \ell)  C_{\ell}^{(1)}\left(1 + \tfrac{2U }{ \delta}\right)}{(1 + \ell)}, \\
R_3(U,\delta) &= - \sum_{\ell=0, 2}^{2 \delta -2} \Big( \frac{f_0 (\delta , \ell)}{12 \delta ^3 (\ell+1) (2 \ell+1)} \big( 4 \delta ^2 \left(3 \ell^2+2 \ell-2\right) C_{\ell}^{(1)}\left(1 + \tfrac{2U }{ \delta}\right) \nonumber \\
&+ 6 \delta  (4 \delta  (\ell-1)-12U) C_{\ell}^{'(1)}\left(1 + \tfrac{2U }{ \delta}\right) -48U (\delta +U)  C_{\ell}^{''(1)}\left(1 + \tfrac{2U }{ \delta}\right) \big) \Big), \nonumber \\
R_2(U,\delta) &= - \sum_{\ell=0, 2}^{2 \delta -2}  \frac{f_0(\delta , \ell)}{24 \delta ^5 (\ell+1) (2 \ell+1)} \Big( 2 \delta ^3  (12 \sqrt{\delta } (2 \ell+1) \tau_2(\delta , \ell)+6 \ell^3-25 \ell^2-138 \ell-32) \nonumber \\
& \times C_{\ell}^{(1)}\left(1 + \tfrac{2U }{ \delta}\right) -4 \delta ^2 \left(\left(3  \ell^2+22 \ell-31\right) 4U+40 \delta  (\ell-1)\right) C_{\ell}^{'(1)}\left(1 + \tfrac{2U }{ \delta}\right) \nonumber \\
&+ 4U(\delta  (8 \delta  (16-3 \ell)+(47-6 \ell) 4U))C_{\ell}^{''(1)}\left(1 + \tfrac{2U }{ \delta}\right)+192 U^2 ( \delta +U)C_{\ell}^{'''(1)}\left(1 + \tfrac{2U }{ \delta}\right) \Big), \nonumber \\
R_1(U,\delta) &= - \sum_{\ell=0, 2}^{2 \delta -2} \frac{f_2(\delta, \ell)}{\delta ^2 (\ell+1) } C_{\ell}^{(1)}\left(1 + \tfrac{2U }{ \delta}\right) \nonumber \\
&+ \sum_{\ell=0, 2}^{2 \delta -2}  \frac{f_0(\delta, \ell)}{96 \delta^7 (\ell+1) (2 \ell+1)}\Big( (192 \sqrt{\delta } (2 \ell+1) \tau_2 (\delta , \ell)+1248 \ell^3+4364 \ell^2+2782 \ell+927) \nonumber \\
&\times \delta^4 C_{\ell}^{(1)}\left(1 + \tfrac{2U }{ \delta}\right) +4 \delta ^3 (48 \sqrt{\delta } (2 \ell+1) U \tau_2(\delta , \ell) \nonumber \\
&+\left(6 \ell^3-37 \ell^2-383 \ell+249\right) 4U-314 \delta  (\ell-1) ) C_{\ell}^{'(1)}\left(1 + \tfrac{2U }{ \delta}\right)     \nonumber \\
&+ 4U \delta ^2 \left(\left(-12 \ell^2-216 \ell+677\right) 4U+4 \delta  (365-104 \ell)\right) C_{\ell}^{''(1)}\left(1 + \tfrac{2U }{ \delta}\right) \nonumber \\
&+ 32 U^2 \delta  (4 \delta  (35-3 \ell)+(45-4 \ell) 4U) C_{\ell}^{'''(1)}\left(1 + \tfrac{2U }{ \delta}\right) \nonumber \\
&+ 768 U^3 (\delta +U) C_{\ell}^{''''(1)}\left(1 + \tfrac{2U }{ \delta}\right)  \Big). \nonumber
\end{align}
We checked that these residues agree with the expressions \eqref{residues} when plugging in the OPE data from section \ref{sec:ope_data}.

Analogously, we can write a two channel dispersion relation in the $U$ and $T$ channels. When we equate it to the previous dispersion relation, and set $U=0$, $T=-S$ and expand around $S=0$ we get
\begin{align}
0  =\approx \alpha^{(0)}_{-1, 1} + \alpha^{(1)}_{-1,1} S + \alpha^{(1)}_{-2,2} S^2 + \ldots\,.
\end{align}
So, we recover the crossing equations $\alpha^{(0)}_{-1, 1}=0$, $\alpha^{(1)}_{-1,1}=0$, $\alpha^{(1)}_{-2,2}=0$, etc.

The next step is to set up the analogue of (\ref{Asymptoticsf0 2}). At large $\delta$ the summand in (\ref{TwoChannelT1}) must behave like $\frac{\log^3 \delta}{ \delta}$ when $U \approx 1$ so as to give rise to a fourth order pole in $U$. The reason for this is that  
\begin{align}
\frac{1}{(U-1)^4} &= \frac{\kappa^4}{6} \lim_{\delta_0 \rightarrow \infty} \int_{\delta_0}^{\infty} d\delta \delta ^{\kappa (U-1)-1} \log^3(\delta)
\end{align}
This reasoning leads to the equation
\begin{align}\label{HugeUgly}
\lim_{\delta \rightarrow \infty}  \sum_{k=1}^{4}  \frac{R_k(1,\delta)}{ \delta^k } = \kappa \frac{\log^3 \delta}{\delta}\,,
\end{align}
where $\kappa$ is some constant, which we cannot determine by this type of argument. (\ref{HugeUgly}) is a generalisation of (\ref{Asymptoticsf0 2}).
Equation (\ref{HugeUgly}) follows purely from bootstrap, and so it can potentially act as a check on the assumption that the $\alpha^{(1)}_{a, b}$'s are single-valued MZVs. In order to check (\ref{HugeUgly}) we need to generate $f_2(\delta, \ell)$ and $\tau_2(\delta, \ell)$ for large values of $\delta$. The amount of nested sums involved unfortunately grows very rapidly, and so we were unable to generate a database of $f_2(\delta, \ell)$ and $\tau_2(\delta, \ell)$ for sufficiently high values of $\delta$ so as to check (\ref{HugeUgly}).

The explicitly known terms in the expression for the residues \eqref{residues} fall off faster than $\log^3 \delta / \delta$, so \eqref{HugeUgly} constrains the functions $h_n(\delta)$ and 
$g_n(\delta)$
\beq
\lim_{\delta \rightarrow \infty}
\frac{1}{\delta^4} \sum\limits_{n=0}^{\delta-1}
\left. \left( h_n(\delta) + g_n(\delta) \left(1- \delta^2 (2+U\partial_U) \right)\right)  R(U+\tfrac{n}{2},\delta-n)\right|_{U=1} = \kappa \frac{\log^3 \delta}{\delta}\,.
\eeq

\section{Leading poles}
\label{app:leading_poles}

The poles of highest order in $A^{(k)}(S,T)$ are universal in the sense that they do not depend on corrections to the OPE data.
For $A^{(1)}(S,T)$ these are the fourth order poles which originate from the terms in \eqref{c2} that are polynomials of degree three in $a$ and $b$.

To understand the structure of these terms better, we can compute the analogous terms at higher orders in $1/\sqrt{\lambda}$ by expanding \eqref{Mdef}.
We compute the terms that produce the leading poles for $k\leq 5$
and find that they have the following form, which we expect to hold for all $k$
\beq
\alpha_{a,b}^{(k)} = 
\sum\limits_{\delta=1}^\infty \sum\limits_{m=0}^b \frac{1}{\delta^{3+2a+3b+k}} \frac{(-1)^k (2a+3b)^{3k}}{6^{k} \Gamma(k+1)}  c^{(0)}_{a,b,m} F_m^{(0)}(\delta) + \text{ subleading poles in }A^{(k)}(S,T)\,.
\eeq
This can be summed over $k$ to give an exponential $\text{exp}\left(- \frac{(2a+3b)^3}{6 \delta \sqrt{\lambda}}\right)$.
We can apply the flat space transform and sum over $a$ and $b$ to get
\bea
2\sum\limits_{k,a,b=0}^\infty \frac{\hat\sigma_2^a \hat\sigma_3^b \alpha_{a,b}^{(k)} }{\lambda^\frac{k}{2}}
={}& 
\sum\limits_{\delta=1}^\infty \frac{1}{\delta^{3}} \text{exp}\left(- \frac{(2 x\partial_x+3 y\partial_y)^3}{6 \delta \sqrt{\lambda}}\right)
 \frac{y+2}{1-x-y} \binom{z+\delta-1}{\delta-1}^2\\
& + \text{ subleading poles}\,.
\eea{summing_alpha_k}
The leading poles arise when the derivatives act on the denominator
\beq
(2 x\partial_x+3 y\partial_y)^{3k} \frac{y+2}{1-x-y}
= \frac{(3k)! S^{3k+1}}{(\delta-S)^{3k+1}} + O\left( \frac{1}{(S-\delta)^{3k}} \right)\,, \qquad \delta = 1,2,\ldots\,.
\eeq
This gives
\begin{align}
2\sum\limits_{k,a,b=0}^\infty \frac{\hat\sigma_2^a \hat\sigma_3^b \alpha_{a,b}^{(k)} }{\lambda^\frac{k}{2}}
={}& \sum\limits_{k=0}^\infty
\frac{1}{\lambda^\frac{k}{2}}
\left( \frac{R^{(k)}_{3k+1} (T,\delta)}{(S-\delta)^{3k+1}} + O\left( \frac{1}{(S-\delta)^{3k}} \right) \right) \,,
\label{summing_alpha_k_poles}
\end{align}
with
\beq
R^{(k)}_{3k+1} (T,\delta) = -\frac{(3k)! \delta^{2k-2}}{6^k k!} 
\frac{\Gamma (T+\delta )^2}{\Gamma (\delta )^2 \Gamma (T+1)^2}\,.
\eeq
The leading order poles can be Borel resummed. We obtain

\begin{equation}
\sum_{k=0} - \frac{(3k)! \delta^{2k-2}}{6^k k!} \frac{1}{\lambda^{k/2} (S-\delta)^{3k+1}} =- \int_0^\infty dt \frac{e^{\frac{\delta ^2 t^3}{6 \sqrt{\lambda } (S-\delta )^3}-t}}{\delta ^2 (S-\delta )} \,.
\end{equation}
The integral can be written in terms of Airy and hypergeometric functions, if desired.

\section{Dimension of the Konishi operator}\label{app:Konishi}

At weak coupling, the dimension of the Konishi operator (which is related to one of the operators exchanged in our correlator as discussed below \eqref{Delta_to_DeltaHat}) was calculated in \cite{Leurent:2013mr}. It is given by
\begin{align}\label{WeakCoupling}
{}&\hat\Delta_{\text{weak}} = 4 + 12 g^2 -48 g^4 + 336 g^6 + 96 g^8 (6 \zeta (3)-15 \zeta (5)-26) \nonumber\\
& -96 g^{10} \left(54 \zeta (3)^2-72 \zeta (3)+90 \zeta (5)-315 \zeta (7)-158\right)  \nonumber \\
&-48 g^{12} \left(432 \zeta (3)^2-3240 \zeta (5) \zeta (3)+5472 \zeta (3)-2340 \zeta (5)-1575 \zeta (7)+10206 \zeta (9)+160\right) \nonumber \\
&+48 g^{14} (2592 \zeta (3)^3-8784 \zeta (3)^2+8568 \zeta (5) \zeta (3)-40320 \zeta (7) \zeta (3)+108960 \zeta (3)-20700 \zeta (5)^2 \nonumber \\
&-4776 \zeta (5)-26145 \zeta (7)-17406 \zeta (9)+152460 \zeta (11)-44480 ) \nonumber \\
&+ 96 g^{16}\big( 20736 \zeta (3)^3+82656 \zeta (3)^2-90 (504 \zeta (5)+721 \zeta (7)-2688 \zeta (9)+9664) \zeta (3) \nonumber \\ 
&+24840 \zeta (5)^2+227799 \zeta (7)+72 \zeta (5) (3220 \zeta (7)-2847)+97164 \zeta (9)-1104246 \zeta (13) \nonumber \\
&+ 566752 -\tfrac{9}{5}\left( 22800 \zeta (3)^2 \zeta (5)-194711 \zeta (11) +792 \zeta^{\text{sv}}(5,3,3) \right)  \big)\,,
\end{align}
where $g=\frac{\sqrt{\lambda}}{4 \pi}$. At order $g^{16}$ we see that $\zeta^{\text{sv}}(5, 3, 3)$ appears! This suggests that $\hat\Delta_{\text{weak}}$ can be written in terms of single valued MZVs only.  

At strong coupling, the analytic prediction is \cite{Gromov:2011bz}
\begin{align}\label{StrongCouplingAnalytics}
\hat\Delta_{\text{strong}} = 2 \lambda^{\frac{1}{4}}  +2 \lambda^{-\frac{1}{4}}+\left(\frac{1}{2}-3 \zeta (3)\right)\lambda^{-\frac{3}{4}} + \left( 6 \zeta (3)+\frac{15 \zeta (5)}{2}+\frac{1}{2} \right)\lambda^{-\frac{5}{4}}\,.
\end{align}
Two more orders were computed numerically in \cite{Hegedus:2016eop}
\begin{align}\label{StrongCouplingNumerics}
\hat\Delta_{\text{strong}} \sim -91.976023725 \lambda^{-\frac{7}{4}} +758.514613 \lambda^{-\frac{9}{4}}\,.
\end{align}
It would be interesting to find a function $\hat\Delta(\lambda)$ such that
\begin{itemize}
\item at large $\lambda$ it can be expanded in powers of $\lambda^{-\frac{1}{4}}$, matches (\ref{StrongCouplingAnalytics}) and (\ref{StrongCouplingNumerics}), and all numerical coefficients are single valued periods,

\item at small $\lambda$ it can be expanded in even powers of $g$, it matches (\ref{WeakCoupling}), and all numerical coefficients are single valued periods.
\end{itemize}

\bibliographystyle{JHEP}
\bibliography{paper2}
\end{document}